\documentclass[3p,,preprint,11pt, twocolumn]{elsarticle}
\makeatletter\if@twocolumn\PassOptionsToPackage{switch}{lineno}\else\fi\makeatother

\usepackage{tabulary,xcolor}
\usepackage{amsfonts,amsmath,amssymb}
\usepackage[T1]{fontenc}
\makeatletter
\let\save@ps@pprintTitle\ps@pprintTitle
\def\ps@pprintTitle{\save@ps@pprintTitle\gdef\@oddfoot{\footnotesize\itshape \null\hfill\today}}
\def\hlinewd#1{%
  \noalign{\ifnum0=`}\fi\hrule \@height #1%
  \futurelet\reserved@a\@xhline}
\def\tbltoprule{\hlinewd{.8pt}\\[-12pt]}
\def\tblbottomrule{\noalign{\vspace*{6pt}}\hline\noalign{\vspace*{2pt}}}
\def\tblmidrule{\noalign{\vspace*{6pt}}\hline\noalign{\vspace*{2pt}}}
\AtBeginDocument{\ifNAT@numbers \biboptions{sort&compress}\fi}
\makeatother

\usepackage{ifluatex}
\ifluatex
\usepackage{fontspec}
\defaultfontfeatures{Ligatures=TeX}
\usepackage[]{unicode-math}
\unimathsetup{math-style=TeX}
\else 
\usepackage[utf8]{inputenc}
\fi 
\ifluatex\else\usepackage{stmaryrd}\fi

\usepackage{url,multirow,morefloats,floatflt,cancel,tfrupee}
\makeatletter

\AtBeginDocument{\@ifpackageloaded{textcomp}{}{\usepackage{textcomp}}}
\makeatother
\usepackage{colortbl}
\usepackage{xcolor}
\usepackage{pifont}
\usepackage[nointegrals]{wasysym}
\makeatletter

\def\mcWidth#1{\csname TY@F#1\endcsname+\tabcolsep}

\def\cAlignHack{\rightskip\@flushglue\leftskip\@flushglue\parindent\z@\parfillskip\z@skip}
\def\rAlignHack{\rightskip\z@skip\leftskip\@flushglue \parindent\z@\parfillskip\z@skip}

\@ifundefined{etal}{}{}

\usepackage{ifxetex}
\ifxetex\else\if@twocolumn\@ifpackageloaded{stfloats}{}{\usepackage{dblfloatfix}}\fi\fi

\AtBeginDocument{
\expandafter\ifx\csname eqalign\endcsname\relax
\def\eqalign#1{\null\vcenter{\def\\{\cr}\openup\jot\m@th
  \ialign{\strut$\displaystyle{##}$\hfil&$\displaystyle{{}##}$\hfil
      \crcr#1\crcr}}\,}
\fi
}

\AtBeginDocument{%
  \@ifpackageloaded{endfloat}%
   {\renewcommand\efloat@iwrite[1]{\immediate\expandafter\protected@write\csname efloat@post#1\endcsname{}}}{\newif\ifefloat@tables}%
}%

\def\BreakURLText#1{\@tfor\brk@tempa:=#1\do{\brk@tempa\hskip0pt}}
\let\lt=<
\let\gt=>
\def\processVert{\ifmmode|\else\textbar\fi}

\@ifundefined{subparagraph}{
\def\subparagraph{\@startsection{paragraph}{5}{2\parindent}{0ex plus 0.1ex minus 0.1ex}%
{0ex}{\normalfont\small\itshape}}%
}{}

\newcommand\role[1]{\unskip}
\newcommand\aucollab[1]{\unskip}
  
\@ifundefined{tsGraphicsScaleX}{\gdef\tsGraphicsScaleX{1}}{}
\@ifundefined{tsGraphicsScaleY}{\gdef\tsGraphicsScaleY{.9}}{}
\def\checkGraphicsWidth{\ifdim\Gin@nat@width>\linewidth
	\tsGraphicsScaleX\linewidth\else\Gin@nat@width\fi}

\def\checkGraphicsHeight{\ifdim\Gin@nat@height>.9\textheight
	\tsGraphicsScaleY\textheight\else\Gin@nat@height\fi}

\def\fixFloatSize#1{}
\let\ts@includegraphics\includegraphics

\def\inlinegraphic[#1]#2{{\edef\@tempa{#1}\edef\baseline@shift{\ifx\@tempa\@empty0\else#1\fi}\edef\tempZ{\the\numexpr(\numexpr(\baseline@shift*\f@size/100))}\protect\raisebox{\tempZ pt}{\ts@includegraphics{#2}}}}

\AtBeginDocument{\def\includegraphics{\@ifnextchar[{\ts@includegraphics}{\ts@includegraphics[width=\checkGraphicsWidth,height=\checkGraphicsHeight,keepaspectratio]}}}

\DeclareMathAlphabet{\mathpzc}{OT1}{pzc}{m}{it}

\def\URL#1#2{\@ifundefined{href}{#2}{\href{#1}{#2}}}

\def\UrlOrds{\do\*\do\-\do\~\do\'\do\"\do\-}%
\g@addto@macro{\UrlBreaks}{\UrlOrds}

\edef\fntEncoding{\f@encoding}

\makeatother

\newif\ifmultipleabstract\multipleabstractfalse%
    \makeatletter
\def\ead{\@ifnextchar[{\@uad}{\@ead}}
\gdef\@ead#1{\bgroup
   \def\_{\string\underscorechar\space}
   \def\{{\string\lbracechar\space}
   \def\textdagger{\string\textdagger\space}
   \def\texttildeapprox{\string\texttildeapprox\space}
   \def~{\hashchar\space}
   \def\}{\string\rbracechar\space}
   \edef\tmp{\the\@eadauthor}
   \immediate\write\@auxout{\string\emailauthor
     {#1}{\expandafter\strip@prefix\meaning\tmp}}
  \egroup
}
\gdef\emailauthor#1#2{\stepcounter{ead}
      \g@addto@macro\@elseads{\raggedright
      \let\corref\@gobble
      \eadsep\texttt{#1} (#2)
      \def\eadsep{\unskip,\space}}
}

\makeatother
  
\usepackage{float}

\begin{document}

\begin{frontmatter}

    \title{
  Understanding the main failure scenarios of subsea blowout preventers systems: An approach through Latent Semantic Analysis    
}
    
\author[]{Gustavo Jorge Martins de Aguiar}
\author[]{Ramon Baptista Narcizo}
\author[]{Rodolfo Cardoso}
\author[]{Iara Tammela}
\author[]{Edwin Benito Mitacc Meza}
\author[]{Danilo Colombo}
\author[]{Luiz Ant{\^o}nio de Oliveira Chaves}
\author[]{Jamile Eleut{\'e}rio Delesposte}
    
\address{
    Federal Fluminense University\unskip, Rua Recife\unskip, Rio das Ostras\unskip, 28895-532\unskip, Rio de Janeiro\unskip, Brazil}

\begin{abstract}
The blowout preventer (BOP) system is one of the most important well safety barriers during the drilling phase because it can prevent the development of blowout events. This paper investigates BOP system's main failures using an LSA-based methodology. A total of 1312 failure records from companies worldwide were collected from the International Association of Drilling Contractors' RAPID-S53 database. The database contains recordings of halted drilling operations due to BOP system's failures and component's function deviations. The main failure scenarios of the components annular preventer, shear rams preventer, compensated chamber solenoid valve, and hydraulic regulators were identified using the proposed methodology. The scenarios contained valuable information about corrective maintenance procedures, such as frequently observed failure modes, detection methods used, suspected causes, and corrective actions. The findings highlighted that the major failures of the components under consideration were leakages caused by damaged elastomeric seals. The majority of the failures were detected during function and pressure tests with the BOP system in the rig. This study provides an alternative safety analysis that contributes to understanding blowout preventer system's critical component failures by applying a methodology based on a well-established text mining technique and analyzing failure records from an international database.
\end{abstract}
    
  \end{frontmatter}

\section{Introduction}
When the pressure in the underground reservoir exceeds the pressure applied by the drilling fluid column, an uncontrolled flow of gas, oil, or other well fluids occurs through the wellbore and into the atmosphere or the sea, depending on whether the operation is subsea or surface\unskip~\cite{1143479:22575717}. This event is known as a blowout.\textbf{\space }Though rare, blowouts are among the most feared and violent accidents in the oil and gas industry, posing a significant threat to assets, human lives, and the environment\unskip~\cite{1143479:22575738}. Blowouts in deepwater oil and gas well exploration and development cause not only massive environmental disasters and property losses but also fatalities\unskip~\cite{1143479:22575719}. It is one of the major events that contribute to the risks of offshore drilling operations\unskip~\cite{1143479:22575747,1143479:22575643}. Blowouts risk alone contributes 40 percent to 50 percent of the total risk of loss of life, environmental impact, and economic value loss for fixed platforms in the North Sea\unskip~\cite{1143479:22575685}. 

The blowout preventer (BOP) system is one of the most important well safety barriers during the drilling phase because it can prevent the development of blowout events by closing and sealing the well\unskip~\cite{1143479:22575638}. Due to the risks of blowout events, blowout preventers are critical to the monitoring and maintenance of well integrity and the safety of the crew and the environment\unskip~\cite{1143479:22575751}. The Gulf of Mexico, the North Sea, offshore West Africa, and offshore Brazil are experiencing increasing oil and gas exploration and production from deepwater locations with water depths exceeding 300 meters\unskip~\cite{1143479:22575700}. Deepwater drilling activities utilize subsea blowout preventers, and the distance between the BOP system and the drilling rig is approximately three kilometers\unskip~\cite{1143479:22575855}. Drilling operations face operational and technological challenges that are becoming increasingly complex due to the extreme variables of the deepwater drilling seabed environment, such as low temperature and high pressure\unskip~\cite{1143479:22575644}. 

A blowout preventer is a group of valves remotely controlled from the rig and serves as the main barrier in well control in the event of a blowout\unskip~\cite{1143479:22575638,1143479:22575766}. While the primary form of control in the drilling phase is hydrostatic pressure, which is counterbalanced by the weight of the drilling mud, the BOP system is considered one of the last resources capable of preventing blowouts\unskip~\cite{1143479:22575767}. Blowout preventers are designed to seal the well during emergency control events and test and training scenarios to guarantee the system's capability to perform the required function\unskip~\cite{1143479:22576192}. Recognized industry regulations and standard requirements have been implemented as guidelines for testing strategies that can be used to detect potential failures and reduce blowout preventer system unavailability\unskip~\cite{1143479:22575645,1143479:22575742,1143479:22575674,1143479:22575716}. Despite their importance in performing the required function, the components are primarily monitored through testing, and numerous functional and pressure tests are required to ensure the safety barriers during operations\unskip~\cite{1143479:22575699}.

The world realized the critical importance of the BOP system after the Macondo well blowout accident at the Deepwater Horizon semi-submersible drilling rig in 2010\unskip~\cite{1143479:22575638,1143479:22575713,1143479:22575751}. The blowout resulted in 11 deaths, 4.9 million barrels of crude oil spilled, and significant pollution and ecological damage\unskip~\cite{1143479:22575751,1143479:22575709}. Aside from the estimated \$42 billion in losses, the event raised concerns about the BOP system and compelled the oil and gas industry to establish new maintenance and operating standards to improve BOP reliability\unskip~\cite{1143479:22575702,1143479:22575713,1143479:22575750}. Following the accident, BOP risk analysis became more critical in oil exploration and production projects, and the public's perception of the risk of offshore drilling increased due to the event's consequences\unskip~\cite{1143479:22575672,1143479:22575670}. However, risk management practices have not kept up with the growing complexity of drilling operations (due to the industry's expansion into deepwater areas), leading to more severe disasters\unskip~\cite{1143479:22575719}.\textbf{\space }

Failures in equipment and components of complex systems are problems that have long demanded engineering attention for different purposes\unskip~\cite{1143479:22575630,1143479:22575725}. Deviations from the operating condition of equipment or systems can cause not only financial damage but also loss of human life and environmental damage\unskip~\cite{1143479:22575725}. In this context, offshore drilling also requires a high level of safety. Offshore drilling rig blowouts and explosions pose significant financial, environmental, and human life risks\unskip~\cite{1143479:22575672,1143479:22575735,1143479:22575746}. Failures in the blowout preventer can affect the availability of the entire production system because of the high risks\unskip~\cite{1143479:22575727}. Making decisions in the face of indications of failure of one of the BOP system's components is one of the industry's most significant challenges\unskip~\cite{1143479:22575739}. 

In many instances, when a BOP's failure is detected, the equipment must be taken out of operation for repair, and the return to operation may take up to a week or more depending on availability on board\unskip~\cite{1143479:22575713}. Furthermore, the withdrawal of the blowout preventer stack and the marine drilling riser system due to failure is one of the most expensive downtime (non-productive time) events in offshore drilling, and millions of dollars that are lost each year due to downtime can be avoided through improvements in BOP system's reliability\unskip~\cite{1143479:22575669}. This event becomes even more costly than suspension due to scheduled maintenance\unskip~\cite{1143479:22575631}. In\unskip~\cite{1143479:22575686}, subsea BOP system's failures in the Gulf of Mexico between 2007 and 2009 were investigated, and the study collected 156 failures that resulted in 13448 hours (560 days) of downtime. Given that any problem or failure that requires the withdrawal of the blowout preventer costs approximately \$1.2 million per day\unskip~\cite{1143479:22575750}, a total of \$672.4 million was lost as a result of the 156 failures.

Conducting BOP system's failure studies is critical in assisting oil and gas companies in identifying which systems and components are prone to failure and, as a result, providing knowledge about failure's scenarios for use in loss prevention decision-making\unskip~\cite{1143479:22575679}. The study of blowout preventer system failures requires a systematic analysis of the BOP system and an understanding of the main critical failures and their consequences if a blowout occurs\unskip~\cite{1143479:22575673}. 

More and more companies in the oil and gas industry are striving to collect data and create databases about their operations to extract knowledge and support future management decisions\unskip~\cite{1143479:22575732}. Failure records are essential because they are used as input for data analytics processes to improve industry safety, performance, and knowledge\unskip~\cite{1143479:22575731}. While text descriptions of the interventions written by the maintenance technicians can be found in a database of maintenance records, maintenance optimization is frequently limited to information about the time of maintenance interventions due to the difficulty of automatically analyzing texts\unskip~\cite{1143479:22575743}. Natural language efforts in maintenance data pose a significant challenge due to its unstructured nature and technical language with particular terms\unskip~\cite{1143479:22575757}.

Few studies have examined maintenance text records to understand equipment maintenance interventions better and improve future decisions. In\unskip~\cite{1143479:22575666}, a semi-supervised conditional random field (CRF)-based information extraction approach is proposed to extracting information entities from bridge inspection reports identifying existing deficiencies, and perform maintenance actions. In\unskip~\cite{1143479:22575721}, a text mining method is used to extract the most common product failure stories and their causes of failure and regression analyses to look into the links between consumer repair experiences and future purchasing behaviors. In\unskip~\cite{1143479:22575743}, a method is proposed to using textual information to identify equipment degradation states by clustering maintenance records and developing a stochastic multi-state degradation model using a Convolutional Neural Network (CNN). In\unskip~\cite{1143479:22575660}, a text mining method is developed to understand defects of a secondary device in an intelligent substation that combines global vectors for word representation (GloVe) and attention-based bidirectional long short-term memory (BiLSTM-Attention). In\unskip~\cite{1143479:22575689},  a text mining method is used to extracting information about failure patterns in building systems and components from CMMS databases on interesting parts of the database containing work orders about component's failures. In\unskip~\cite{1143479:22575701}, a hybrid natural language processing (hybrid-NLP) algorithm is used to extract entities that represent electrical equipment. In\unskip~\cite{1143479:22575695}, a sustainable fault diagnosis model based on imbalanced text mining and natural language processing technology is used to extract fault feature words from field fault data. In\unskip~\cite{1143479:22575648}, a method is developed to identifying root cause factors by extracting root cause text from unstructured data using a keyword extraction method. In\unskip~\cite{1143479:22575677}, a deep learning method is proposed to processing a power grid malfunction report that combines text data mining oriented recurrent neural networks (RNN) with long short-term memory (LSTM). In\unskip~\cite{1143479:22575697}, a text mining method is used to extract additional information reports and used an integrated spatial-temporal approach, namely the Geographically and Temporally Weighted Ordered Logisgression (GTWOLR) to model the natural gas pipeline incident reports. In\unskip~\cite{1143479:22575615}, a new methodology based on a series of steps is proposed to preprocess and decompose the service history to identify relevant words and sentences that distinguish an unhealthy wind turbine from a healthy one.

This study aims to identify and analyze the most common failure scenarios of the following components of the BOP system: annular, shear rams, regulator, and compensated chamber solenoid valve. Furthermore, as well, to determine how the failures were detected and which maintenance actions were taken. In order to accomplish this, latent semantic analysis (LSA) was used to generate concepts that represent topics based on BOP corrective maintenance text records from an international database known as RAPID-S53, which is derived from the BOP Reliability Joint Industry Project managed by the International Association of Drilling Contractors (IADC). The LSA was used in this paper because it can retrieve multiple conceptual topics based on documents with similar contexts, resulting in topic groupings of documents and terms\unskip~\cite{1143479:22575754,1143479:22575737}. Given the records' context, these concepts can be interpreted as failure scenarios incorporating domain expert knowledge. Understanding the scenarios is critical for acting to improve industry safety by reducing accident risks and increasing equipment availability.

LSA has proven to be a valuable tool for analyzing reviews, products feedbacks, maintenance interventions reports, and other textual records capable of assisting decision-making\unskip~\cite{1143479:22575617}. Customers were segmented and analyzed by\unskip~\cite{1143479:22575715} using a combination of group RFM analysis and probabilistic latent semantic analysis models, and the results indicated that the developed approach provides insight and captures a wide range of customer preferences. In\unskip~\cite{1143479:22575654}, Latent Semantic Analysis, Text2Vec, and Doc2Vec techniques are used to analyze data from the Parkinson's Progression Markers Initiative (PPMI). In\unskip~\cite{1143479:22575646}, the main accident topics in a database of railroad equipment accident descriptions maintained by the Federal Railroad Administration in the United States are identified using LSA and LDA. In\unskip~\cite{1143479:22575754}, groups of contextually similar terms from future-oriented data sources, including experts' and the general public's concerns about drone technology, are extracted using the LSA. In\unskip~\cite{1143479:22575753}, the LSA is used to extract essential topics in a large group of paper abstracts from the field of Multiple Criteria Decision Making (MCDM), and they were able to identify principle categories and significant themes contained in the text.

The proposed research's main contribution is that the information gathered could assist in the development of maintenance planning actions, reliability studies, and support in identifying critical conditions for the well drilling system's shutdown decision. Furthermore, text mining results provide a rich additional source of data for future predictive analytics workflows, and leveraging unstructured data resources allows for more accurate predictive models\unskip~\cite{1143479:22575731}.

The paper is structured as follows. Section 2 discusses the theoretical background and relevant literature on accidents in the oil and gas industry and the components and functions of blowout preventer systems. Section 3 describes the research methodology, which includes data collection and the use of LSA. Sections 4 and 5 present the research findings and the discussion. Section 6 presents the findings' conclusions and implications. Finally, Section 7 discusses the research's limitations as well as suggestions for further studies.
    
\section{Theoretical background}

\subsection{Blowouts in offshore drilling}The demand for oil and gas sources has increased in recent decades due to economic growth, and oil and gas remain the primary energy resources\unskip~\cite{1143479:22575672}. Drilling is a crucial part of the oil and gas industry, and its success will be determined by its ability to improve its processes' operational reliability and availability significantly\unskip~\cite{1143479:22575626}. Deepwater drilling life cycle phases are well planning, drilling, and completion, with the drilling phase encompassing many activities such as drilling, running casing, cementing, circulation, fluid displacement, and clean-up\unskip~\cite{1143479:22575655}. Given that deepwater drilling entails complex operations that must be completed in short periods and that errors can cost tens of millions of dollars, balancing risks, schedules, and budgets is a difficult task\unskip~\cite{1143479:22575655,1143479:22575623}.

Because of the significant financial investment required and the fact that drilling wells are hazardous operations, safety is a top priority, and the activities are strictly regulated through regulations and laws\unskip~\cite{1143479:22575661}. Accidents during the drilling phase generally halt production and harm the image of companies involved, so measures to reduce the frequency and severity of accidents are critical\unskip~\cite{1143479:22575633}. Furthermore, as the industry approaches reservoirs at deeper water depths and in more complex geological formations, drilling operations become more complex, resulting in increased risks\unskip~\cite{1143479:22575719}. The well control operation aims to keep the well integrity by addressing the procedures to be followed when formation fluids begin to flow into the wellbore\unskip~\cite{1143479:22575722,1143479:22575738}. Blowout is the uncontrolled flow of formation fluids to the surface, and it is one of the major events that contribute to the risks associated with offshore drilling\unskip~\cite{1143479:22575643}.

A blowout occurs when a kick (the flow of formation fluid into the wellbore)~is not discovered early enough, or when safety barriers in the well, such as blowout preventers, fail to seal\unskip~\cite{1143479:22575738}. When the formation pressure overcomes the pressure exerted by the~fluid column, such as drilling fluid,~a blowout occurs\unskip~\cite{1143479:22575742}. Blowouts are an expensive and feared operational hazard in offshore drilling, causing delays as well as fires, explosions, casualties, asset damage, and environmental damage\unskip~\cite{1143479:22575685}.

Because of the severe consequences of deepwater blowouts, scholars have conducted extensive research on them\unskip~\cite{1143479:22575670}. April 20, 2010, well blowout, also known as the Deepwater Horizon accident, prompted questions about the safety of deepwater drilling\unskip~\cite{1143479:22575751}. The gas exploded and caught fire on the rig's deck, killing eleven workers, and the blowout caused oil to flow for two months from the damaged well\unskip~\cite{1143479:22575751}. This environmental disaster affects local economies, sensitive coastlines, and wildlife throughout the Gulf region and still\unskip~\cite{1143479:22575687,1143479:22575650,1143479:22575692}. On August 16, 1984, a blowout occurred on the Enchova Central oil rig in Brazil. The majority of the workers on the platform were safely rescued, but 42 people died due to a failure in the lifeboat lowering system, and six of them died while jumping from a height of 30 to 40 meters into the water\unskip~\cite{1143479:22575712}. A second blowout occurred on April 24, 1988, when the BOP could not seal the well and attempts to prevent the event failed. The platform was destroyed after a drill pipe was forced out of the well and struck one of the platform's legs, causing sparks to ignite gas from the blowout\unskip~\cite{1143479:22575633}. The operators of the Montara H1 oil and gas well lost control of the well on August 21, 2009, resulting in a blowout. There were no serious injuries or deaths due to the incident, but uncontrolled hydrocarbons spilled into the atmosphere for 75 days\unskip~\cite{1143479:22575661}. On October 21, 1982, on Eugene Island Block 361, well-control was lost in the Gulf of Mexico, and a significant blowout and fire occurred. One or more shallow gas sands with slightly abnormal pressures charged the gas flow, which flowed outside the drill pipe and upward through the annular preventer, resulting in a full-scale gas blowout\unskip~\cite{1143479:22575734}. 

According to\unskip~\cite{1143479:22575633} analysis from data collected since 1956, blowouts are historically the most common type of accident in all world regions. In North America, 38 accidents involving blowouts were recorded, accounting for 38\% of all accident types in the region. The same pattern was observed in Europe and for North Sea operations, with 9 (28\%) blowouts recorded. Blowouts were also the most common accident type in South America, Asia, and Africa, accounting, respectively, for 29 (82.8\%), 12 (46.2\%), and 8 (57.2\%). This data emphasizes the dominance of blowouts over other types of accidents during well drilling operations.

\subsection{Blowout preventers system}A BOP is an electro-hydraulic system used to seal, control and monitor oil and gas wells\unskip~\cite{1143479:22575638,1143479:22575742}. It was built to handle high pressures and prevent blowouts by sealing the well with an annular preventer and shear rams preventers\unskip~\cite{1143479:22575722}. The two main subsystems of the subsea BOP system are the BOP stack and the control system \unskip~\cite{1143479:22575645}.  A blowout preventer stack is constituted of one or two annular preventers, three to six ram preventers, two connectors (one connecting the BOP to the wellhead connector and the other connecting the lower marine riser package to the BOP), and four to ten choke and kill valves, according to\unskip~\cite{1143479:22575716}. Drilling companies use several annular preventers and ram preventers as system redundancies to improve the reliability of the BOP stack\unskip~\cite{1143479:22575638}. Figure~\ref{figure-4fcb74d53c3f43e3a97c7c51f1b2ff7a}  shows a typical configuration of a subsea BOP system.

\bgroup
\fixFloatSize{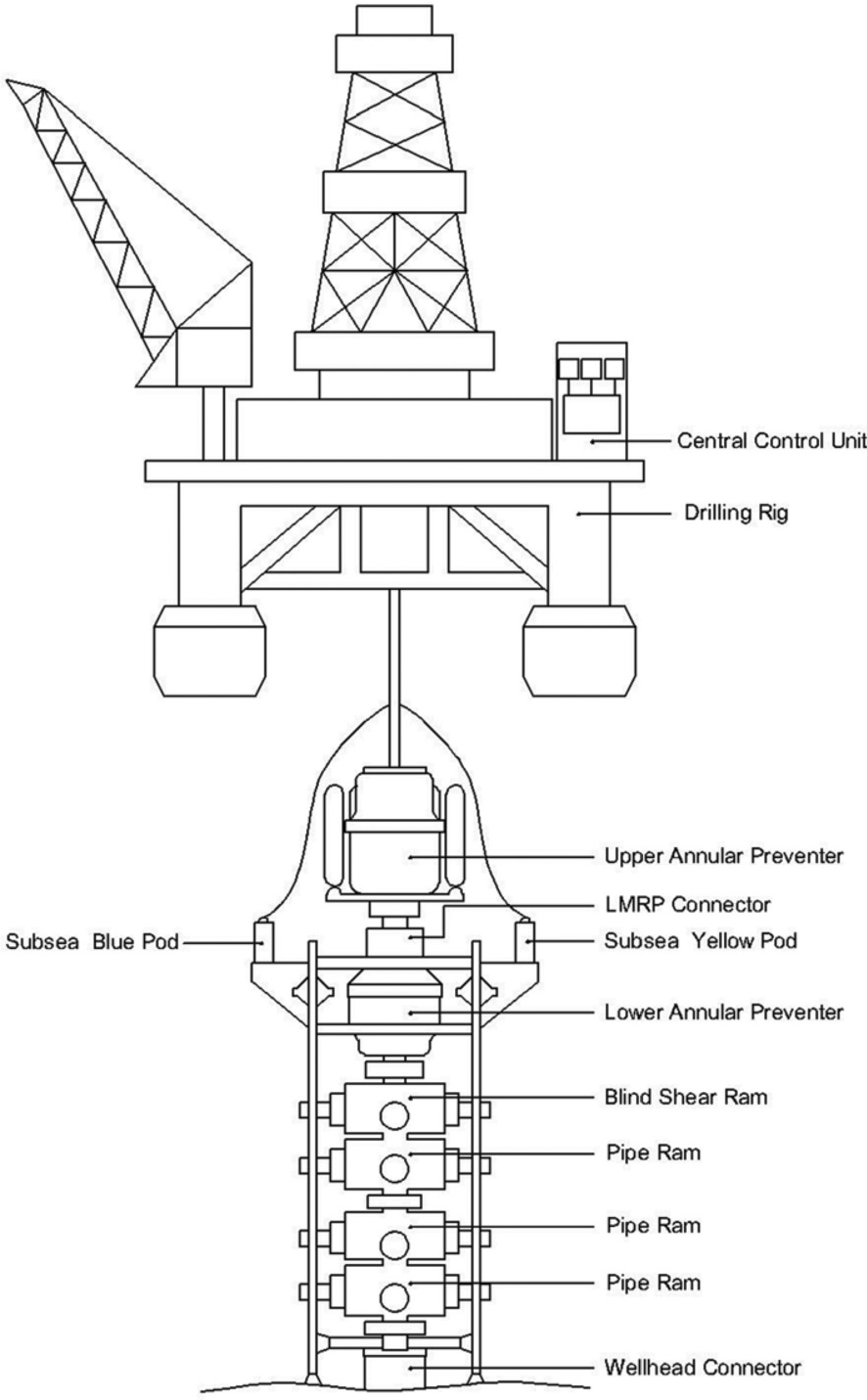}
\begin{figure}[!htbp]
\centering \makeatletter\IfFileExists{images/risa1918-fig-0001-m.jpg}{\includegraphics[width=.93\linewidth]{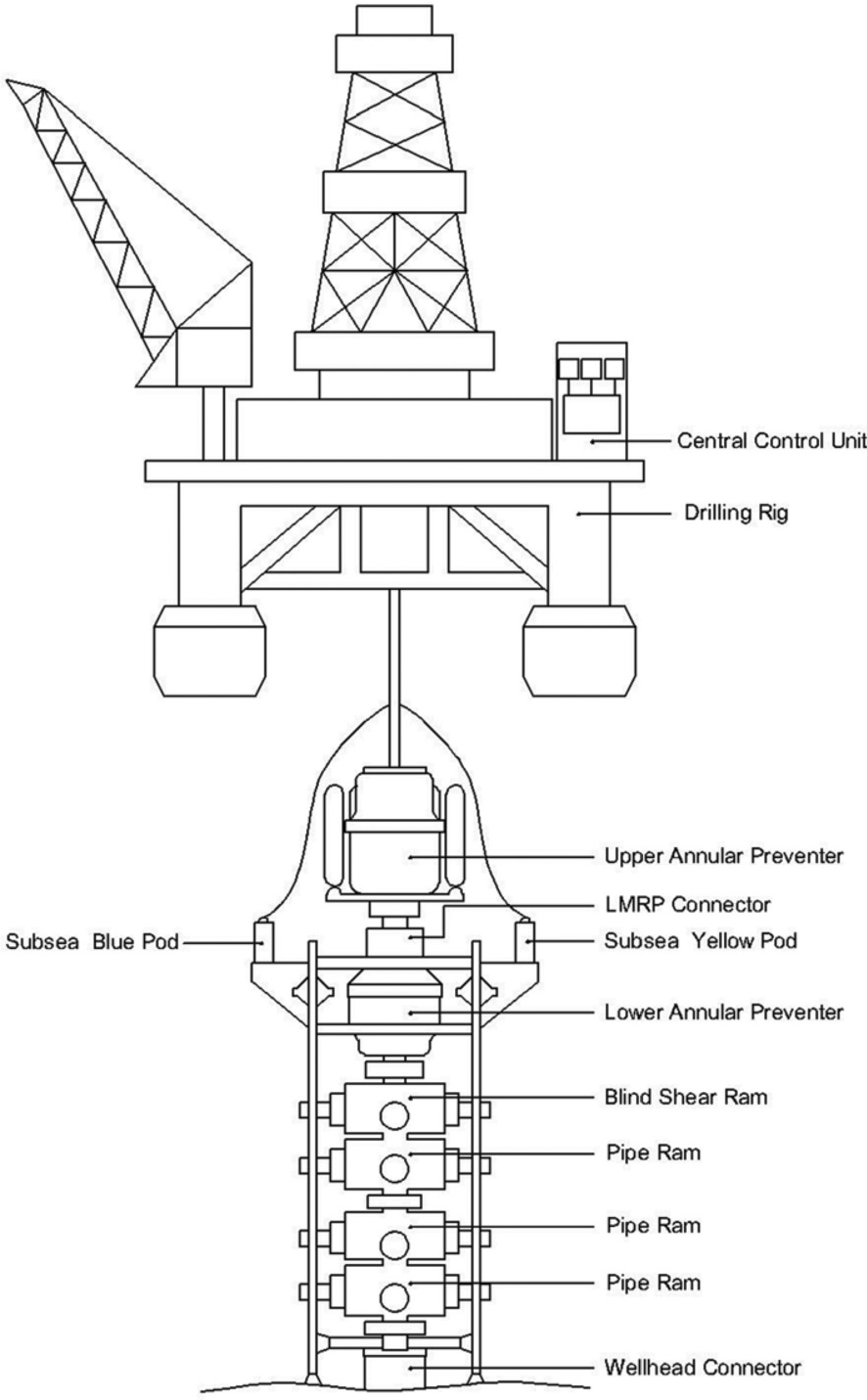}}{}
\makeatother 
\caption{{Typical configuration of a subsea BOP stack\unskip~\protect\cite{1143479:22575613}.}}
\label{figure-4fcb74d53c3f43e3a97c7c51f1b2ff7a}
\end{figure}
\egroup
One of the most critical barriers capable of preventing blowouts during deepwater drilling operations is the annular blowout preventer\unskip~\cite{1143479:22575711}. An annular preventer is a blowout preventer that seals between the tube and the wellbore or an open hole using a shaped sealing elastomeric component\unskip~\cite{1143479:22575742}. The annular preventer effectively maintains a seal around the drill pipe even as it rotates during drilling, but it is not as effective as ram preventers at sealing on an open hole, and it must be capable of closing the wellbore to comply with regulations\unskip~\cite{1143479:22575638}.

The steadily increasing wall thickness, because drilling ultradeep wells places significant demands on the drill string, required to absorb high tensile loads, has exceeded the capacity of some shear rams to shear drill pipe successfully in some cases\unskip~\cite{1143479:22575628}. Pipe rams, blind shear rams (BSR), and casing shear rams (CSR) are the three main types of ram preventers. Pipe rams close around a drill pipe, restricting flow between the drill pipe and the wellbore\unskip~\cite{1143479:22575638}. While blind shear rams are designed to cut the drill string as the rams close off the well, casing shear rams should be able to cut the casing as well, and they should be able to effectively seal the hole against the flow of oil and gas in emergencies such as potential blowouts\unskip~\cite{1143479:22575645}. Among the various devices on the BOP, blind shearing rams serve as the last line of defense, and two BSRs are used in some deepwater BOP to increase the chances of cutting the drill pipe and sealing the well\unskip~\cite{1143479:22575698,1143479:22575720}.

Although there are two types of subsea BOP control systems, hydraulic and multiplexed (MUX), the MUX system is the more recent and widely used because as exploration depth increased, problems with the reaction time of umbilicals used in the hydraulic control system were observed, and hydraulic lines controlling the pilot valves were replaced with separate electric cables operating the compensated chamber solenoid valves (CCSV)\unskip~\cite{1143479:22575716,1143479:22575679}. Most subsea blowout preventers used for deepwater drilling are similar to those used for shallow-water drilling, but the BOP is controlled by a multiplex control system, which reduces the time it takes for the BOP functions to activate\unskip~\cite{1143479:22575720}.

The blowout preventer MUX control system comprises two subsystems: an electrical system and a hydraulic system that includes components such as pumps, valves, accumulators, fluid storage and mixing equipment, and a manifold\unskip~\cite{1143479:22575638,1143479:22575645}. The drilling rig's surface control components include three computers in the central control unit (CCU), the driller's station, and the toolpusher's station, as well as three programmable logic controllers (PLC)\unskip~\cite{1143479:22575647,1143479:22575745}. The CCU is the brain behind the subsea BOP control system, and it uses the PLCs to send commands from surface control stations and panels to the subsea control point of distribution (POD)\unskip~\cite{1143479:22575647}. Six fiber optic repeaters complete the communications between the CCU and the blue and yellow subsea electronics modules (SEM) via two fully independent subsea umbilical cables made up of optical fibers and electrical wires to transmit signals and power\unskip~\cite{1143479:22575745}. The blue and yellow SEMs are designed to decode surface signals, and their complete independence allows for a fully redundant system for controlling all subsea functions and communicating with the CCU\unskip~\cite{1143479:22575662}. The decoded signal activates an electrical solenoid valve in the subsea POD, which sends a hydraulic pilot signal to the appropriate function hydraulic valve, also known as subplate-mounted valves (SPM)\unskip~\cite{1143479:22575662}.

Control points of distribution (PODs) are essential to the performance of a subsea control system.~Each POD contains the compensated chamber solenoid valves, pressure transducers, subplate-mounted valves, pressure regulators, flowmeters, and hydraulic accumulators\unskip~\cite{1143479:22575647,1143479:22575642}. PODs are identical and can be mounted in either the blue or yellow location\unskip~\cite{1143479:22575742}. Decoded signals operate the PODs from the SEMs, but only one of them receives hydraulic fluid for performing BOP functions, making the operation of the other POD ineffective\unskip~\cite{1143479:22575730}. The CCSV then sends a hydraulic signal to the corresponding SPM, which is actuated and sends pressurized hydraulic fluid to the BOP system's function\unskip~\cite{1143479:22575662}.

Hydraulic regulators, which reduce the pressure in the supplied fluid through the rigid hydraulic line before entering the POD, are another important component of the BOP control system\unskip~\cite{1143479:22575742,1143479:22575716}. The regulator's spools control sizes of orifices of supply, output, and vent ports according to the pressure downstream and upstream the valve, but there are two types of pressure regulators valves present in the BOP control system\unskip~\cite{1143479:22575755}. The first type is a manual regulating valve known as manual koomey regulators (MKR) and has the desired outlet pressure set on the surface using the adjustment nut before it is put into operation in the BOP system\unskip~\cite{1143479:22575653}\textbf{. }The other is remotely adjustable regulators known as hydraulic koomey regulators (HKR), which can be regulated remotely from the surface via the BOP system's control panel\unskip~\cite{1143479:22575755}. 

Each POD contains an accumulator used in the MUX system to store hydraulic fluid to be used in case of hydraulic flow demands and the loss of power supply to the pumps\unskip~\cite{1143479:22575716,1143479:22575653}. The accumulators in the BOP stack should provide enough volume and pressure of an available hydraulic fluid to actuate the specified well control equipment and enough residual pressure to maintain sealing capability\unskip~\cite{1143479:22575742,1143479:22575705}. To better understand how accumulators operate, consider an order to close the BOP ram. As presented before, the signal will be transmitted from the central control unit (CCU) to the SEM for decoding and then to the POD. The specific compensated chamber solenoid valve will then open to performing the function, triggering the SPM valve to change position and allowing high-pressure fluid stored in the accumulator to pass through, closing the ram\unskip~\cite{1143479:22575730}. 

BOP components are mainly monitored through tests, and two of the most critical tests performed on the BOP are the pressure test and the function test\unskip~\cite{1143479:22575699}. A function test is an operation of a BOP system's component that aims to verify its intended operation (that it can do what it is intended to do) and must be performed at least once a week\unskip~\cite{1143479:22575742}. Only the ability to perform the function of a BOP is tested when it is function tested, which is insufficient to ensure the safety of drilling operations because the capability of sealing the wellbore is not guaranteed\unskip~\cite{1143479:22575643}. Pressure testing the BOP entails evaluating both the capability to act the BOP function and seal off a~pressure, it must also be performed regularly, with no more than 21 days between tests, and it consists of two tests: the low-pressure test and the high-pressure test\unskip~\cite{1143479:22575742}.
    
\section{Methodology}
The method was divided into three major phases: initial procedures, LSA procedures, and post-LSA procedures. Each major phase was made up of stages and smaller steps, and the first phase, initial procedures, consisted of collecting failure records and preparing data. The second phase, LSA procedures, consists of preprocessing, which prepares the failure descriptions for analysis, and LSA processing of the preprocessed records. The third phase is the post-LSA procedures, which begin with creating failure scenarios using the concepts-terms and document-concepts matrices, followed by the final validation of the MFS matrix by domain experts. The method was applied to BOP corrective maintenance text records from the RAPID-S53 international database, managed by the International Association of Drilling Contractors (IADC). Figure~\ref{f-afc6847dc1bf} shows an overview of the methodology of this research.
\bgroup
\fixFloatSize{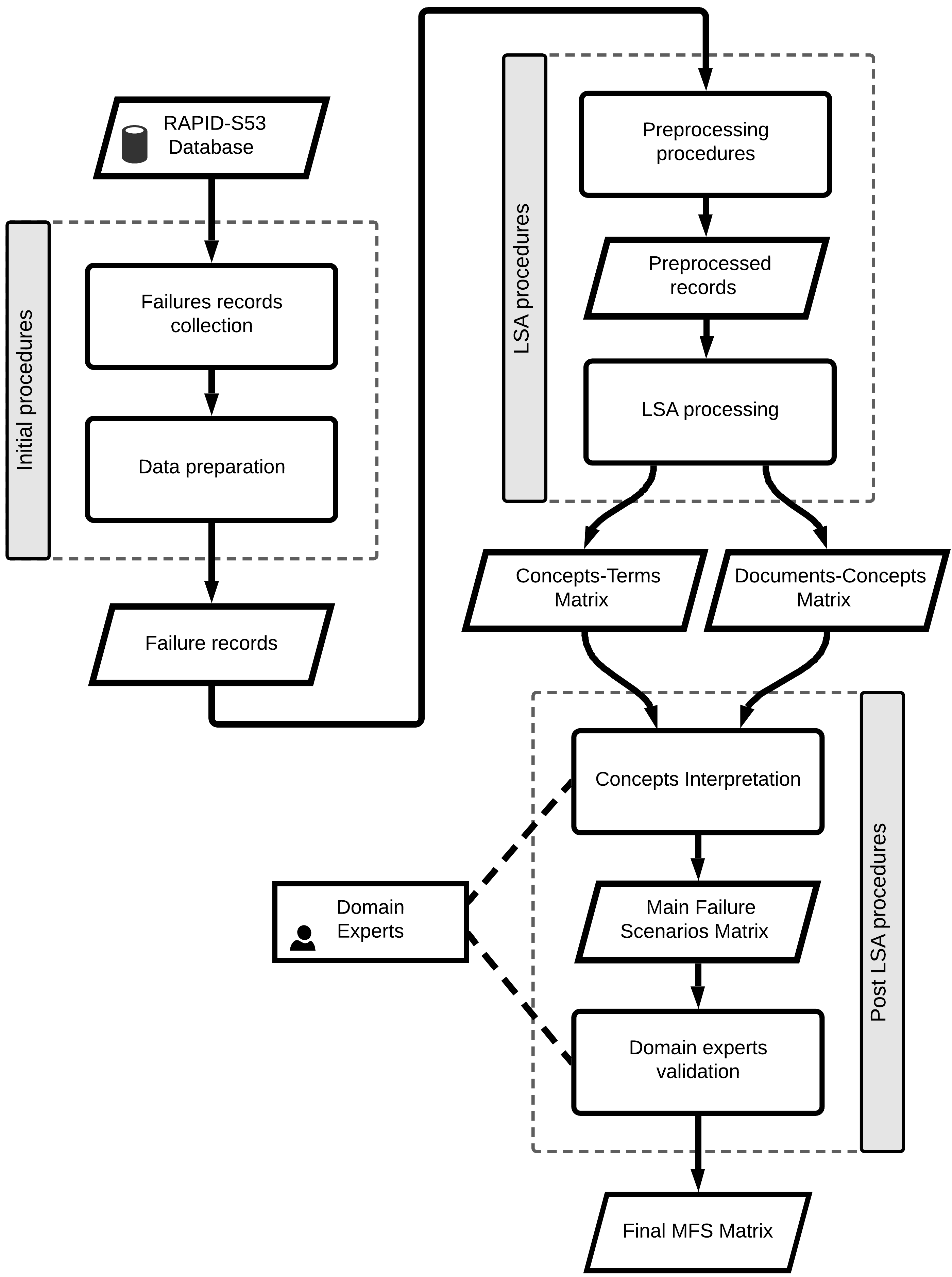}
\begin{figure}[!htbp]
\centering \makeatletter\IfFileExists{images/f47ec9c9-7f66-4c92-9cbe-e4c1132b4214.png}{\includegraphics[width=.97\linewidth]{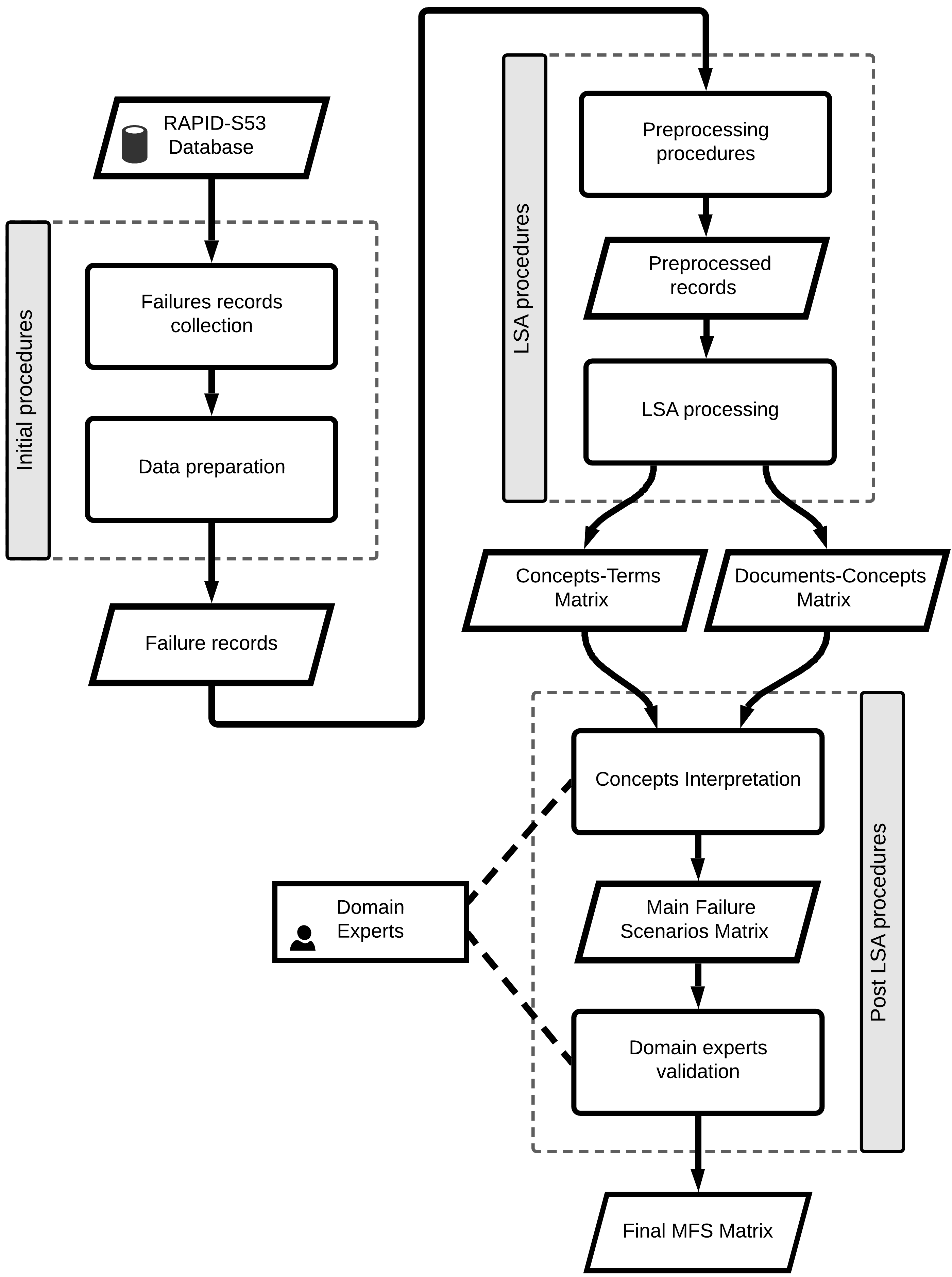}}{}
\makeatother 
\caption{{Overview of proposed methodology.}}
\label{f-afc6847dc1bf}
\end{figure}
\egroup

\subsection{Data collection}The RAPID-S53 database used in this research results from the BOP Reliability Joint Industry Project, in which oil and gas exploration companies, drilling contractors, and original equipment manufacturers (OEM) from all over the world participated. The database includes global records of deviations in equipment functions and drilling operations interruptions associated with the BOP system, and it is maintained by the International Association of Drilling Contractors (IADC). The main objective of this database is to provide a large amount of data that individual companies can use to improve the reliability and efficiency of well control equipment.

There are guidelines for the register of events in the RAPID-S53 database. It is necessary to generate a report for each event in which a component of the well control equipment was considered to be malfunctioning. If no immediate physical corrective action is required to be applied directly to a component in order for it to operate as designed, the event does not need to be reported. However, events during maintenance and testing should continue to be reported, but only defects (failures) and not any event related to preventive maintenance, such as a component preventive replacement\unskip~\cite{1143479:22575684}. 

The data was collected from RAPID-S53 to a worksheet. There are records of failure events from January 2012 to November 2018 from 26 different companies, 6380 registers. Although the registers recorded information related to the failure event, such as the amount of usage at the time of failure, hours of repair time, hours of non-productive time, this research focused on the event description to identify and understand the main failure scenarios of the BOP system. The failure texts are 448 terms long on average, corresponding to two or three paragraphs explaining the failure observed, the components and parts affected, detection methods, and corrective actions.

\subsection{Data preparation}Given the complexity of the blowout preventer system, which is made up of several subsystems with numerous components and parts, it was necessary to work with subsets of the registers to improve the ability of the concepts generated by the LSA approach to describe the main failure scenarios for some components perceived more critical for the BOP system.

In\unskip~\cite{1143479:22575643}, data on subsea BOPs on the Gulf of Mexico's outer continental shelf for ten months, from July 1997 to May 1998, is gathered and the 117 failures recorded resulted in 3638 hours of downtime (non-productive time). The ram preventers account for 41\% of the downtime, with a total time loss of 1505 hours, the primary control system accounts for 28\% (1021 hours), and the annular preventer accounts for 9\% (337 hours) of the downtime caused by BOP's failures. In\unskip~\cite{1143479:22575686}, a reliability study that observed subsea BOP's failures in the Gulf of Mexico is presented, and the data comprises 156 failures registered and 13448 hours of non-productive time from 2007 to 2009. Failure events caused by the BOP control system, such as regulator's failure, solenoid valve's failure, and control fluid leak, were the primary contributors, accounting for 35\% of the unplanned non-productive time, with a total time loss of 4712 hours. Annular preventers and ram preventers are the second and fourth most significant sources of downtime events in\unskip~\cite{1143479:22575686}, accounting for 17\% (2345 hours) and 13\% (1766 hours).

Key components of the blowout preventer system were chosen for analysis using the data presented. The data was segmented based on the components identified in the failure records. The latent semantic analysis was performed on textual descriptions of failures of the following components: annular preventers, shear ram preventers, compensated chamber solenoid valves, and hydraulic regulators. It is critical to note that the database does not differentiate between blind shear ram and casing shear ram. However, it was expected that the LSA applied to the failure event description field would allow for the distinction of the two components' failure scenarios. Similarly, the component indication as regulator is used in the RAPID-S53 database to record failures of both the hydraulic koomey regulator (HKR) and the manual koomey regulator (MKR).

The RAPID-S53 was cut into smaller datasets according to the components, and the data associated with the components selected was gathered in specific worksheets. The tokenization of terms (unigrams, bigrams, and trigrams) was then performed with a minimum frequency of 2.5 percent of the total records for each component worksheet to form an initial bag-of-words (BoW). These BoWs were analyzed and used to assist in creating a dictionary of terms containing bigrams and trigrams that are common in a database containing technical texts such as the RAPID-S53. The purpose is that terms such as annular element, soak test, seal plate, vent port, and weep hole, which used to appear separately, will appear in the concepts with the character underline joining the two or three words that make up the term, facilitating concept interpretation. This list was also used to create a synonym dictionary. Figure~\ref{figure-fab8cd211a1342a0a3d65d07bc6f38a3} illustrates the relationship between this study's latent semantic analysis and data preparation.

\bgroup
\fixFloatSize{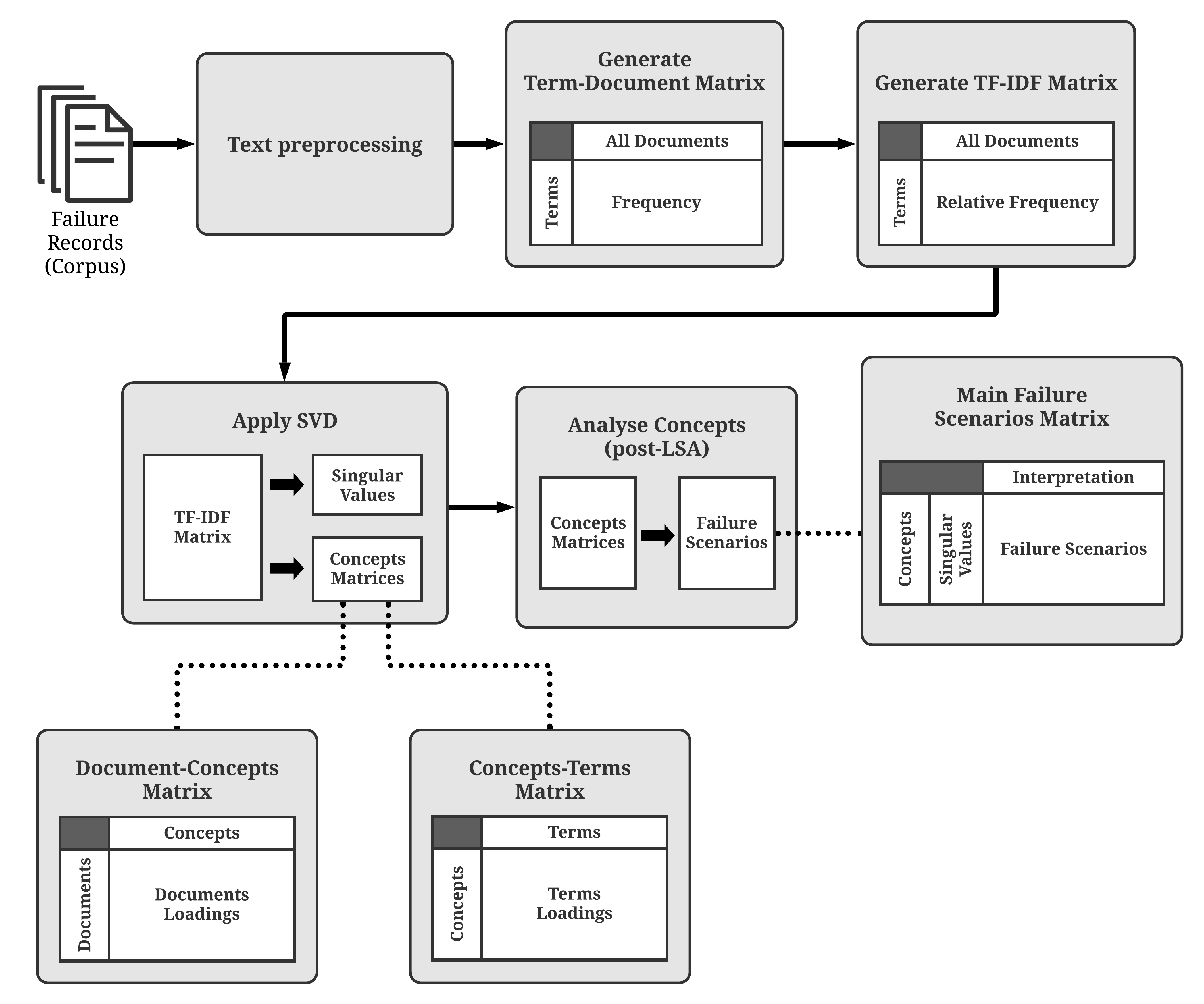}
\begin{figure}[!htbp]
\centering \makeatletter\IfFileExists{images/artigo-sabesp-cxf3pia-de-pxe1gina-1.png}{\includegraphics[width=.95\linewidth]{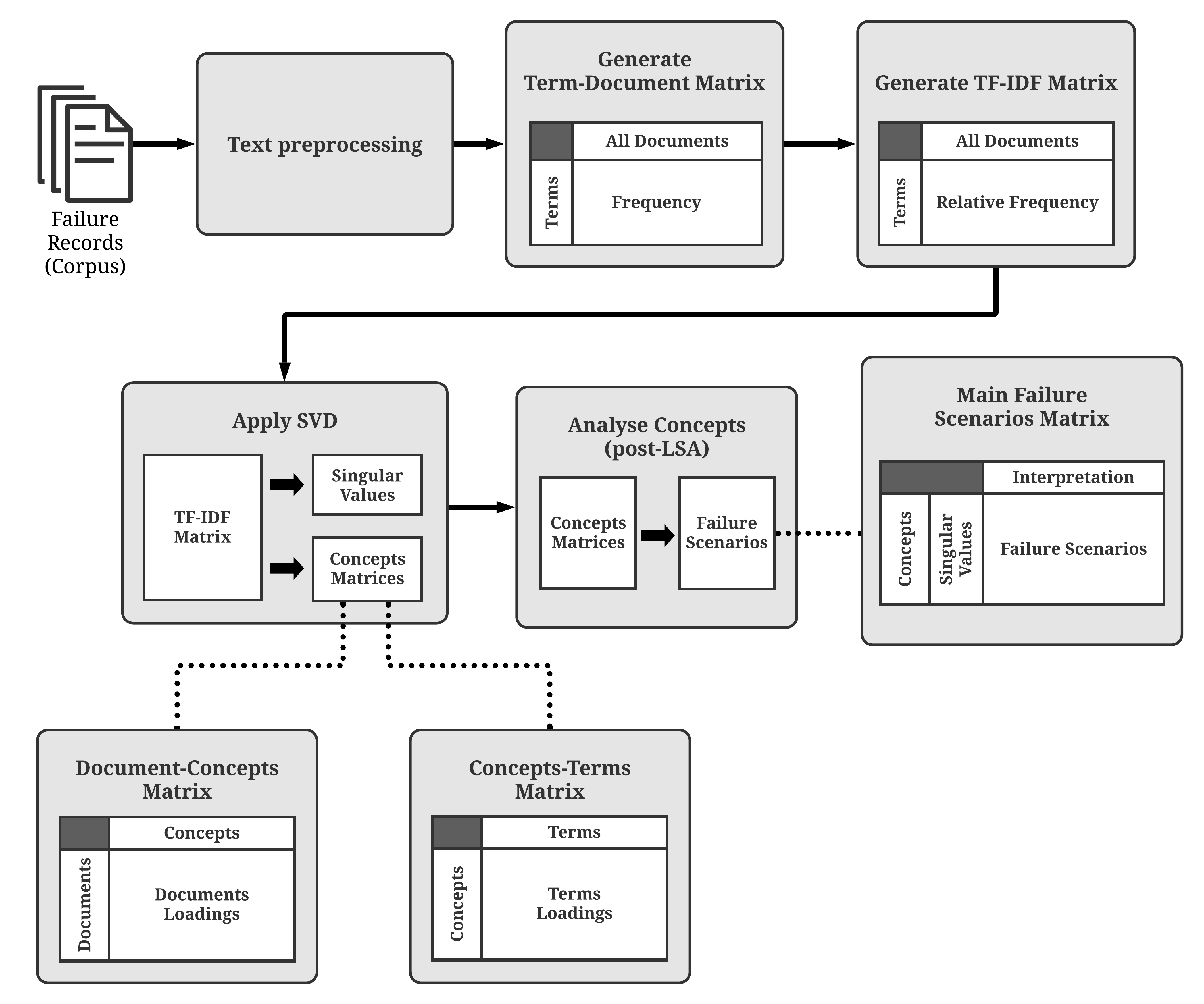}}{}
\makeatother 
\caption{{Overview of the research initial procedures.}}
\label{figure-fab8cd211a1342a0a3d65d07bc6f38a3}
\end{figure}
\egroup
Python, a programming language increasingly being used in academic research, was used in this study\unskip~\cite{1143479:22575656}, was used in this study. Another tool used that has become popular in the data science area is the Jupyter Notebook. The NumPy, Matplotlib, Seaborn, Scikit-learn, Pandas, Pandas-Profiling, NLTK, and SciPy libraries were used in this research\unskip~\cite{1143479:22575680,1143479:22575641,1143479:22575706,1143479:22575724,1143479:22575637,1143479:22575718,1143479:22575694}. 

The specific worksheets correspond to a total of 1312 failure records and 6565 hours of downtime. Considering the previously stated costs of downtime, the failures documented in the worksheets resulted in a monetary loss of more than \$300 million. While annular component failures account for 247 records, contributing to 2778 hours of non-productive time, shear rams' failures account for 310 records, contributing to 1706 hours of downtime. Failures of the regulator and CCSV, both components of the BOP control system, correspond to 421 failures (1121 hours of non-productive time) and 334 failures (960 hours of non-productive time), respectively.

\subsection{Latent Semantic Analysis}Many models have been developed in recent decades to understand the use of words found in textual documents, which is a significant challenge due to the various contexts in which words can be used\unskip~\cite{1143479:22575691}. Because a text is an unstructured form of information with high complexity, and because it is composed of many words or terms (such as bigrams and trigrams) combined to form different ideas throughout the text, reducing the number of terms by removing words that are not relevant to the ideas presented in the text is essential to making the analysis possible\unskip~\cite{1143479:22575627,1143479:22575658,1143479:22575748}. Working with thousands of dimensions can have a negative impact on textual mining results, and reducing the number of dimensions can improve accuracy and efficiency without sacrificing significant aspects of the data\unskip~\cite{1143479:22575726}. The latent semantic analysis (LSA), first introduced as an information retrieval technique\unskip~\cite{1143479:22575736,1143479:22575693} and later defined as knowledge acquisition, induction, and representation theory\unskip~\cite{1143479:22575651}, is a model that can retrieve~topics and ideas~from texts records~with similar contexts\unskip~\cite{1143479:22575754}.

LSA is a well-known mathematical approach for extracting and presenting textual data in a semantic structure\unskip~\cite{1143479:22575754,1143479:22575736}. The mathematical model is based on the idea that words with similar meanings are more likely to occur in similar contexts\unskip~\cite{1143479:22575646}. It acquires semantic knowledge by utilizing word associations found in training documents\unskip~\cite{1143479:22575736}. As a result, the system's interpretation of a word is determined by how the training corpus uses language and how words are associated in the training corpus\unskip~\cite{1143479:22575691}. The model generates semantic representations for words by analyzing the statistical pattern of joint occurrence of words across the training corpus\unskip~\cite{1143479:22575646}. The semantic space constructed from corpus documents contains semantic vectors, also known as concepts or topics. Each of them has a specific value associated with each word in the set of documents, and this value can be interpreted as that word's contribution to the formation of the specific concept\unskip~\cite{1143479:22575693}. Thus, latent semantic analysis can examine the relationships and similarities between documents and the terms (the composition of one or more words) contained within those documents and identify the ideas (topics) presented in texts\unskip~\cite{1143479:22575651}.

LSA consists of three major steps, according to\unskip~\cite{1143479:22575754} and\unskip~\cite{1143479:22575753}. The first step is to create a term frequency (TF) matrix, in which each line represents a word or term, and each column represents a document or context, and individual cell entries contain the frequency with which a term occurs in a document\unskip~\cite{1143479:22575678}. Some studies consider using the preprocessing procedures as the first step to getting better results\unskip~\cite{1143479:22575753}. The frequency of terms is then transformed to form a matrix of terms and documents with transformed values known as term frequency-inverse document frequency (TF-IDF), reflecting how important a word is to a document in a collection\unskip~\cite{1143479:22575617}. Finally, to reduce the matrix's dimensionality was used the singular-value decomposition (SVD), which is an eigenvector decomposition and factor analysis technique\unskip~\cite{1143479:22575736}. This division into three major steps is quite similar to the division presented by\unskip~\cite{1143479:22575678}, which divides into four major steps, the first three of which are identical to those presented and the last of which is associated with information retrieval through vector similarity.

Preprocessing steps, in general, involve filtering out documents of interest to the analyst and eliminating words and terms that are irrelevant\unskip~\cite{1143479:22575681}. Several studies emphasize the importance of preprocessing procedures in text mining research to improve techniques used for information retrieval and topics modeling\unskip~\cite{1143479:22575627,1143479:22575658,1143479:22575748}. Removal of stopwords and stemming are the most commonly used techniques for preprocessing textual data\unskip~\cite{1143479:22575748}.\textbf{\space }Other essential preprocessing techniques include removing unnecessary punctuation, converting letters to lowercase, and lemmatization\unskip~\cite{1143479:22575627}. 

Tokenization, the first activity of the text preprocessing stage in this study, transforms these texts into vectors composed of all of the words and terms contained within them\unskip~\cite{1143479:22575658}. It was possible to identify the essential terms that will be considered in the preprocessing stage and to build a dictionary of terms and synonyms by using initial experimental tokenization. Tokenization was very important for the current research because many technical terms and expressions, such as component names and maintenance procedures, were found frequently in RAPID-S53 database failure records. In this study, unnecessary punctuation was removed, letters were converted to lowercase, and stopwords were removed from the registers in this order. Following that, the texts were lemmatized with WordNet PoS tagging and the WordNet lemmatizer\unskip~\cite{1143479:22575649,1143479:22575625}. According to some studies, lemmatization, despite having a higher computational cost, is more precise than stemmization and produces better results\unskip~\cite{1143479:22575667,1143479:22575620}. Another reason for using lemmatization rather than stemming in this study is that stemming is more likely to produce incorrect or non-existent terms, which is exacerbated by the highly technical nature of the texts describing the BOP system's failures.

Following the preprocessing procedures is the stage of constructing the matrix of terms and documents and calculating the frequency of each term presented in the dictionary built for each failure record. The frequency value is then transformed to construct the TF-IDF matrix, which reduces the influence of words proportionally to their occurrence, given that these words do not significantly contribute to the understanding of the various topics in the records\unskip~\cite{1143479:22575639}. According to\unskip~\cite{1143479:22575678}, there are various methods for calculating the entries of the TF-IDF matrix, and in this study, it was calculated as
\let\saveeqnno\theequation
\let\savefrac\frac
\def\dispfrac{\displaystyle\savefrac}
\begin{eqnarray}
\let\frac\dispfrac
\gdef\theequation{1}
\let\theHequation\theequation
\label{disp-formula-group-b0dfb04b705c4852b20ab25e84e1f237}
\begin{array}{@{}l}w_{ij}={tf}_{ij}\cdot{idf}_i\end{array}
\end{eqnarray}
\global\let\theequation\saveeqnno
\addtocounter{equation}{-1}\ignorespaces 
where, 
\let\saveeqnno\theequation
\let\savefrac\frac
\def\dispfrac{\displaystyle\savefrac}
\begin{eqnarray}
\let\frac\dispfrac
\gdef\theequation{2}
\let\theHequation\theequation
\label{disp-formula-group-cab3f02a49884f918bfd895e9ea8b99c}
\begin{array}{@{}l}idf_i\hphantom{,}={{log\;}{\left(\frac{1+N}{1+n_i}\right)}}+1\end{array}
\end{eqnarray}
\global\let\theequation\saveeqnno
\addtocounter{equation}{-1}\ignorespaces 
$N $ is the total number of records in the corpus, and $n_{i\;} $ is the number of records containing the term indicated by index $i $. The ${idf}_i $ indicates the rarity of occurrence of the term indicated by index $i $ in document $j $, and the higher the value obtained, the rarer it is. The euclidean norm is then applied to the resulting TF-IDF vectors for each document. As a result, the final TF-IDF vectors are denoted as $v_j $ and calculated as follows.
\let\saveeqnno\theequation
\let\savefrac\frac
\def\dispfrac{\displaystyle\savefrac}
\begin{eqnarray}
\let\frac\dispfrac
\gdef\theequation{3}
\let\theHequation\theequation
\label{disp-formula-group-47e81c8da1924289b7a7b3214d6475ba}
\begin{array}{@{}l}v_j=\frac{w_j}{\sqrt{{w_{1j}}^{2}+{w_{2j}}^{2}+\dots+{w_{nj}}^{2}}}\end{array}
\end{eqnarray}
\global\let\theequation\saveeqnno
\addtocounter{equation}{-1}\ignorespaces 
The ability of the singular values decomposition technique to reduce the dimensions of large volumes of data to a more manageable number without losing a significant amount of information from the original variables is why LSA uses it\unskip~\cite{1143479:22575639}. The SVD satisfies the requirement of working with the TF-IDF matrix, which contains a large volume of data\unskip~\cite{1143479:22575736}. 

Mathematically, SVD decomposes the terms and documents matrix or the TF-IDF matrix, $A_{t\times d} $\textbf{, }into the product of three other matrices: $G_{t\times m} $, an orthogonal matrix with m representing the dimensionality of the data, $S_{m\times m} $, a diagonal matrix with single values ordered in descending order, and $D_{d\times m} $, a transpose of the column orthogonal matrix\unskip~\cite{1143479:22575693,1143479:22575736,1143479:22575651}. That is to say,
\let\saveeqnno\theequation
\let\savefrac\frac
\def\dispfrac{\displaystyle\savefrac}
\begin{eqnarray}
\let\frac\dispfrac
\gdef\theequation{4}
\let\theHequation\theequation
\label{dfg-b3b3b770392b}
\begin{array}{@{}l}A_{t\times d}=G_{t\times m}\cdot S_{m\times m} \cdot {{(D}_{d\times m})}^{T}\end{array}
\end{eqnarray}
\global\let\theequation\saveeqnno
\addtocounter{equation}{-1}\ignorespaces 
Where $t $ denotes the number of terms and $d $ denotes the number of documents in the corpus. The matrices are truncated in an arbitrary number of concepts, denoted as $k $ , to remove the noise in the original matrix and thus extract the semantic relations of the documents\unskip~\cite{1143479:22575729}. The SVD result is the best k-dimensional approximation to the original matrix in terms of the least square error\unskip~\cite{1143479:22575678}. 

In the same latent semantic space created by singular values decomposition, each term and document is represented as a k-dimensional vector\unskip~\cite{1143479:22575678}. Each $n $ latent semantic concept, in the interval $I= \left[1,2,\dots, k\right] $, is associated with a set of values, known as loadings, for terms and documents. So, the SVD generates two matrices of concepts loadings, one for the words and one for the documents. These terms and document loadings can be used to interpret or label each concept\unskip~\cite{1143479:22575754}. Figure~\ref{f-d1c81c6961b5} illustrates the LSA stages and outputs.

\bgroup
\fixFloatSize{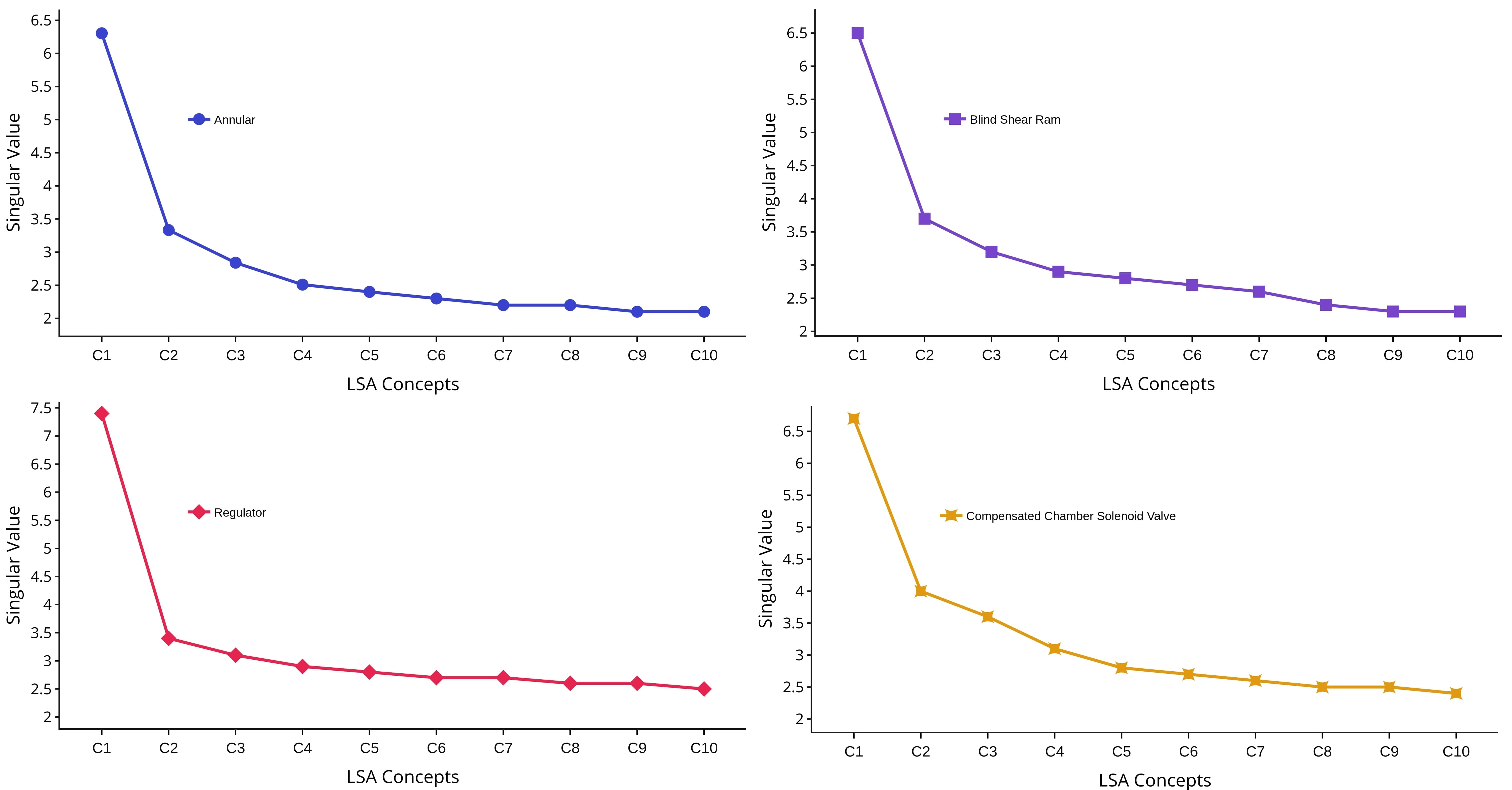}
\begin{figure*}[!htbp]
\centering \makeatletter\IfFileExists{images/merge_from_ofoct-1.png}{\includegraphics[width=.96\linewidth]{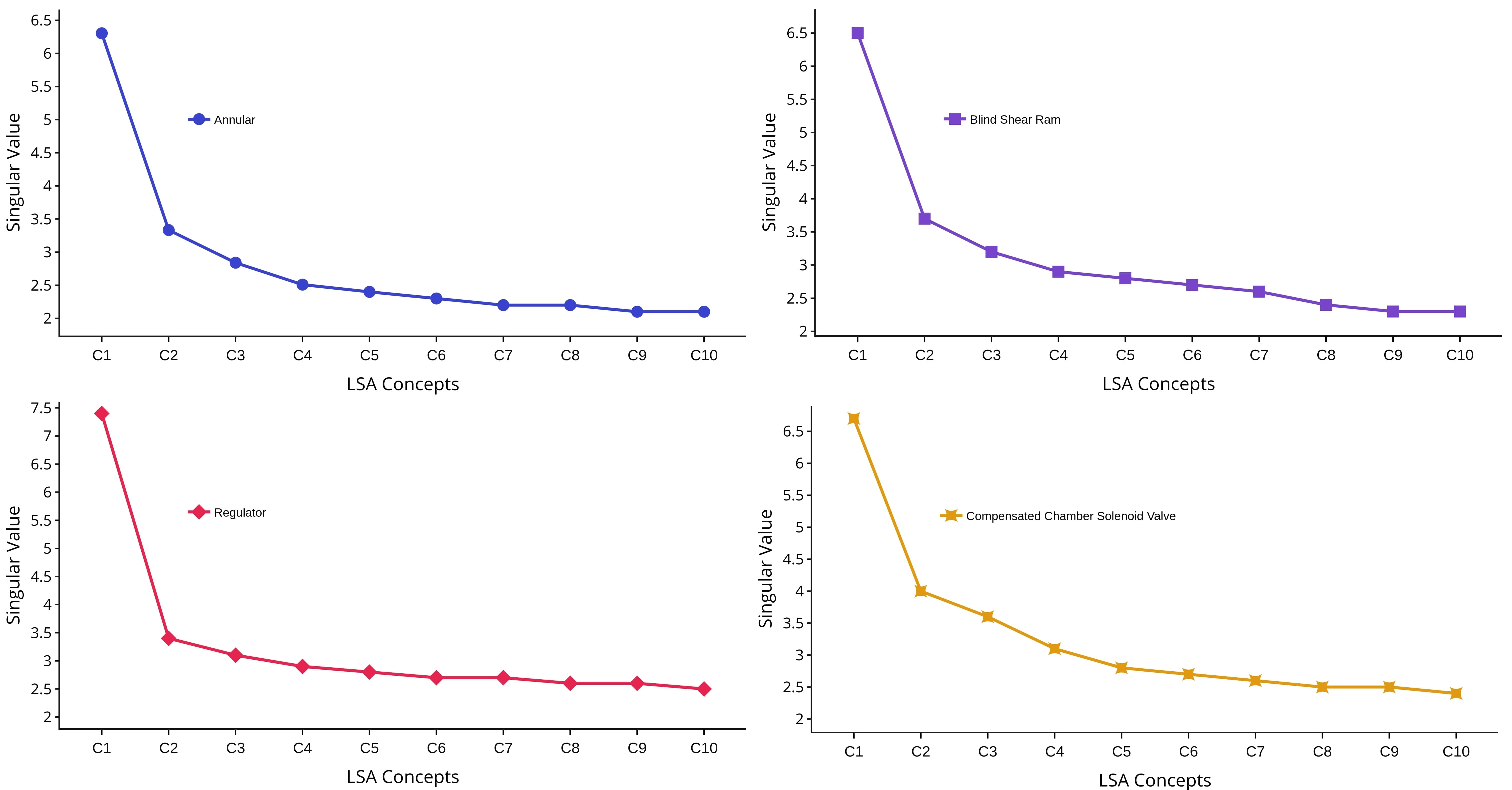}}{}
\makeatother 
\caption{{Overview of LSA and post-LSA procedures and outputs. }}
\label{f-d1c81c6961b5}
\end{figure*}
\egroup
The optimal number of concepts is an open problem in science, and analyzing the graph of the singular values of the concepts produced is one of the most common methods for determining this number\unskip~\cite{1143479:22575639}. The logic goes that the higher the singular value associated with the concept, the better it can explain the data variance\unskip~\cite{1143479:22575744}. In the context of failure records, concepts with higher singular values almost certainly define the components' main failure scenarios (which appeared more frequently in the records). Visualizing the elbow in the graph or when there are accelerating decreasing returns is one possible criterion for defining the concepts under consideration\unskip~\cite{1143479:22575639,1143479:22575671}. Several studies have investigated the terms that contribute the most to concepts (via term loadings) in conjunction with the singular values technique to uncover topics hidden in textual records or documents\unskip~\cite{1143479:22575703,1143479:22575682}. Given the impossibility of analyzing all of the concepts generated by the SVD, using singular values as a cut-off rule to define the number of concepts that should be analyzed is an interesting solution.

Constructing scenarios involves concept interpretation using the term loadings of the Concept-Term (CT) matrix and the documents loadings of the Document-Concept (DC) matrix\unskip~\cite{1143479:22575737}. The high-loading terms and documents were analyzed to identify failure scenarios, which are composed of information regarding corrective maintenance operations. Defining the appropriate term and document loading threshold value is critical and can impact the interpretation process, but there are no established methods for determining the value\unskip~\cite{1143479:22575754}. According to\unskip~\cite{1143479:22575664}\textbf{,} \textbf{\space }researchers must empirically determine the appropriate loading threshold values for each scenario based on coherence. Each concept may have a different threshold value for its terms or documents, but high loading values are required to ensure the scenario is reasonably concrete\unskip~\cite{1143479:22575754}. 

Given that the corpus used in this study was composed of failure records texts, the generated concepts' high loading terms were highly technical. The majority of these terms were component names, detection methods, observed failures, and corrective actions. So, the domain experts responsible for interpreting the concepts and constructing the failure scenarios did a simple classification of the terms, which can support the construction of failure scenarios stage as an initial step of the interpretation process. Figure~\ref{f-331bd337e8b3} shows an example of a high loading terms classification for a theoretical concept. The classification is not necessary for interpreting the concepts, but it does help with interpretation.
\bgroup
\fixFloatSize{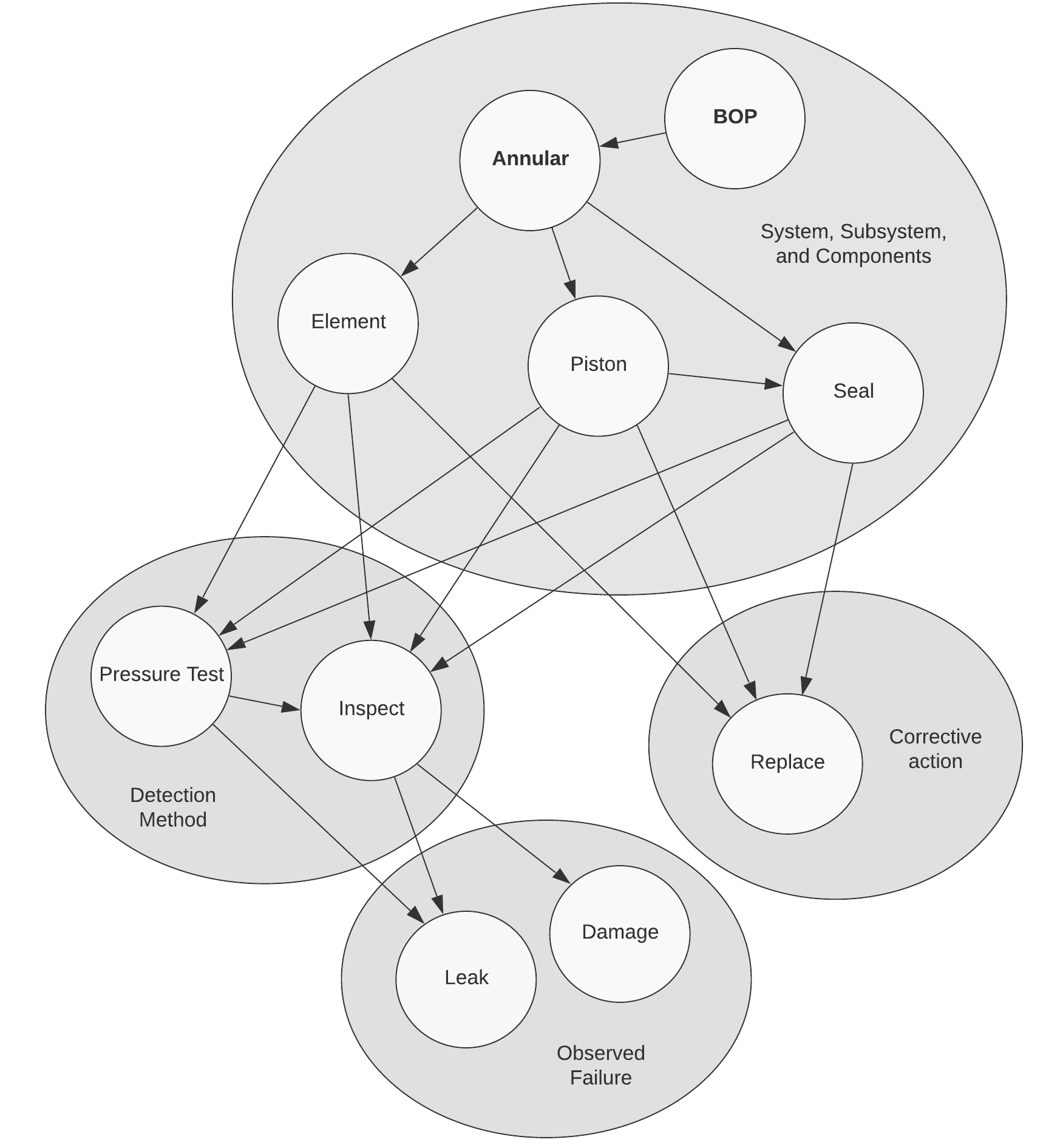}
\begin{figure}[!htbp]
\centering \makeatletter\IfFileExists{images/905e5b5b-17d4-41d0-b4fa-337324cd75e5.png}{\includegraphics[width=.88\linewidth]{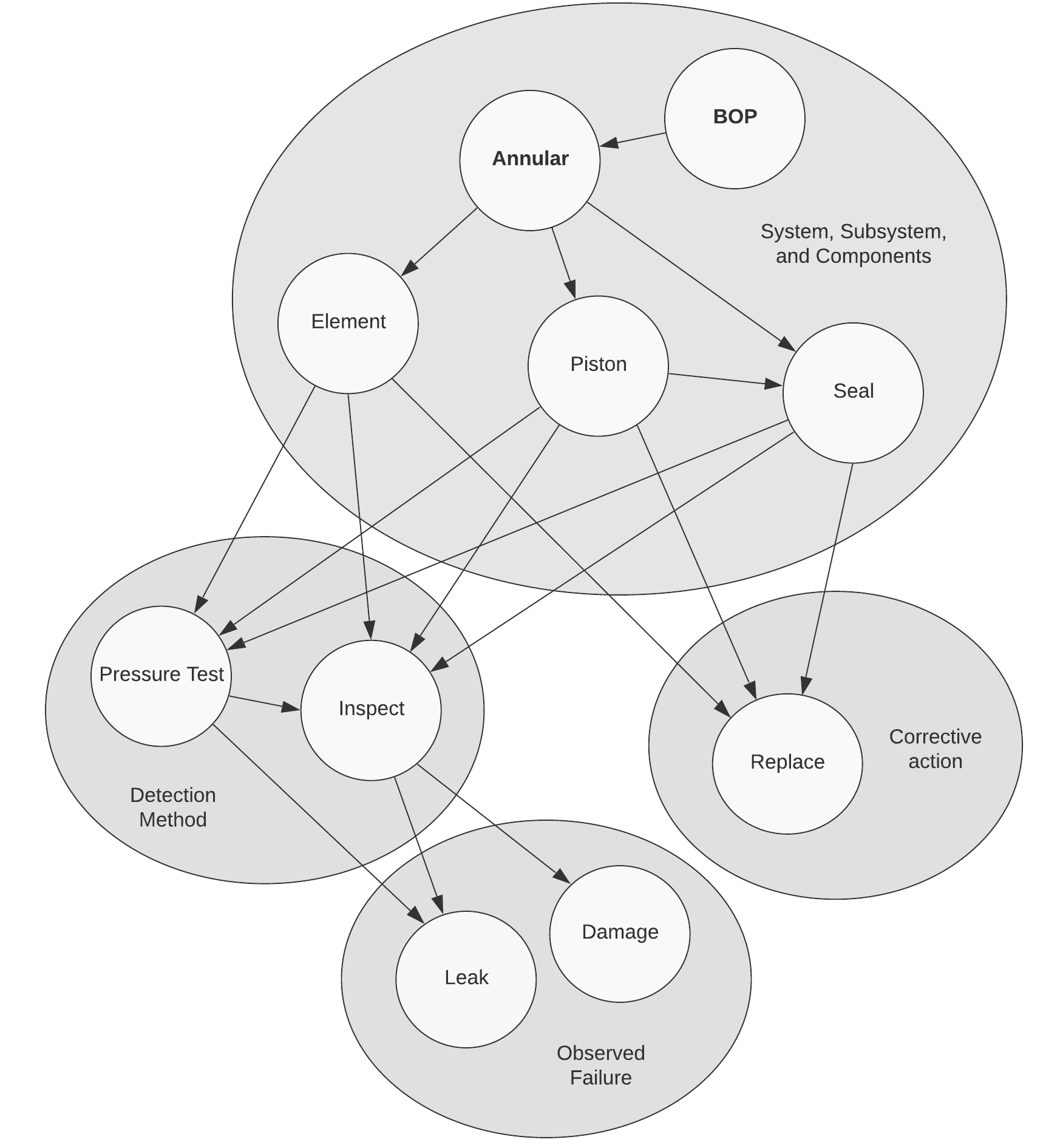}}{}
\makeatother 
\caption{{Example of terms classification of a theoretical concept.}}
\label{f-331bd337e8b3}
\end{figure}
\egroup

\section{Results}
To extract the main failure scenarios, the maximum number of concepts, $k $, was empirically set as ten since this number of concepts is higher than the number of main failure scenarios expected by the authors for each BOP component. It was then used the singular values elbow graph combined with the high loading terms and documents contextual analysis to determine the number of concepts to be analyzed by domain experts. Figure~\ref{f-9e58bd89f06f} shows singular values graphs for each BOP component to the ten concepts with higher singular values.

\bgroup
\fixFloatSize{images/7e14b802-165a-43d8-9873-407afd7ee5f1-umerge_from_ofoct-1.png}
\begin{figure*}[!htbp]
\centering \makeatletter\IfFileExists{images/7e14b802-165a-43d8-9873-407afd7ee5f1-umerge_from_ofoct-1.png}{\includegraphics[width=.90\linewidth]{images/7e14b802-165a-43d8-9873-407afd7ee5f1-umerge_from_ofoct-1.png}}{}
\makeatother 
\caption{{Singular values graphs}}
\label{f-9e58bd89f06f}
\end{figure*}
\egroup
Following the analyses, eight concepts associated with BOP stack components were chosen, four annular preventer's concepts and four shear ram preventer's concepts. As with the BOP stack's components, eight concepts related to BOP control system' components have been selected, five regulator's concepts, and three compensated chamber solenoid valve's concepts. When analyzing the concepts, it was evaluated that a maximum of twenty-five terms was sufficient for understanding each concept's failure scenario. All concepts are given names (denoted by the combination of the component name abbreviation and the number of the order in which it was generated, for example, AC1 for the annular preventer's concept one), and their highest loading terms are reported in Table~\ref{table-wrap-a4649719866f48108d595167e716a1b3} and Table~\ref{tw-a70108a235cc}.

The concepts were independently interpreted by four experts of BOP system's functions, failures, and maintenance operations based on the CT matrix terms loadings of the terms that contribute the most to each concept and the DC matrix highly related failure records, which have high loadings. The results were then compared, and the interpretation was similar for the majority of the scenarios constructed. Some minor disagreements were solved through discussion. The following failure scenarios represent the concepts' final interpretation.

\begin{table*}[t!]
\caption{{BOP Stack components related concepts and their highest loading terms} }
\label{table-wrap-a4649719866f48108d595167e716a1b3}
\def\arraystretch{1}
\ignorespaces 
\centering 
\begin{tabulary}{\linewidth}{p{\dimexpr.10520000000000001\linewidth-2\tabcolsep}p{\dimexpr.10979999999999995\linewidth-2\tabcolsep}p{\dimexpr.785\linewidth-2\tabcolsep}}
\tbltoprule Concept & Singular Values & Terms that contribute the most to each concept\\
\tblmidrule 
\multicolumn{3}{p{\dimexpr(1\linewidth-2\tabcolsep)}}{\textbf{Annular's failures concepts}}\\
 AC1 &
   6.304 &
  seal; annular; test; pressure; leak; pressure\_test; element; maintenance; BOP; piston; new; description; fail; upper\_annular; inspect; lower\_annular; root\_cause; open; replace; instal; observe; close; damage.\\
 AC2 &
   3.333 &
  element; fail; test; annular\_element; root\_cause; pressure\_test; maintenance; description; BOP; rubber; upper\_annular\_element; tool; drill\_pipe; attempt; cycle; joint; pull; pass; protrude; hold; closure; packer.\\
 AC3 &
   2.840 &
  piston; element; score; maintenance; new; description; replace; adapter\_ring; inspect; damage; pit; annular\_piston; cause; root\_cause; change; surface; sent; rubber; disassembly; annular\_element; area; bore; replacement; seal\_area; swarf.\\
 AC4 &
   2.509 &
  seal; lower\_annular; roll; wellbore; upper; cap; o-ring; mud; housing; adapter; replace; instal; fluid; pull; rubber; weep\_hole; packer; element; anti\_extrusion; remove; leak; polypak\_seal; adapter\_ring. \\
\multicolumn{3}{p{\dimexpr(1\linewidth-2\tabcolsep)}}{\textbf{Shear ram's failures concepts}}\\
 SRC1 &
   6.485 &
  seal; pressure; test; bonnet; leak; blind\_shear\_ram; door; BOP; operator; open; maintenance; inspect; pressure\_test; fail; close; o-ring; ram; new; description; damage; remove; hinge; replace.\\
 SRC2 &
   3.717 &
  bolt; blade; inspect; upper; crack; block; ram\_block; low; maintenance; non\_destructive\_test; shear\_ram; fail; pressure\_test; blind\_shear\_ram; rubber; description; lateral\_seal; torque; shear; end\_of\_well.\\
 SRC3 &
   3.186 &
  pressure; pressure\_test; ram; fail; low; rubber; shear\_ram; hold; test; BOP; blind\_shear\_ram; prior; attempt; wellbore; complete; packer; high; surface; change; initial; cavity; good; cycle.\\
 SRC4 &
   2.927 &
  door; hinge; aft; assembly; piston; forward; damage; pressure; year; body; pressure\_test; close\_chamber; remove; notice; crew; poslock; pit; chamber\_test; plug; shear\_ram; cavity; fwd; lock.\\
\tblbottomrule 
\end{tabulary}\par 
\end{table*}
\textbf{Scenario } \textbf{AC1}: The annular's leakage failure is the subject of this scenario. Aside from the leakage failure, the records highly associated with this scenario show that scratches on the annular piston were observed during inspections. Some records described them as light scratches, while others were more specific, stating that they did not exceed an acceptable depth defined by the original equipment manufacturer (OEM). The most affected components are the head seals, chamber seals, piston seals, and annular sealing element (packer element). The leaks were caused by seal cuts, pinching, and wear, according to the records. Because the seals are made of elastomeric material (rubber), seal damage and wear can result in fluid leaks due to leakage paths. There appears to be no difference in the likelihood of failure occurrence for both lower and upper annular preventers. Failures were detected through surface pressure tests (the term surface refers to the fact that the tests were performed with the BOP in the rig). According to the records highly connected to this scenario, after the leakage was noticed, the annular was disassembled at the surface to investigate and identify the cause of the incident. The corrective action is the installation of a new component during maintenance. The observed failure may affect the capability of the BOP system to seal the well in an emergency.

\textbf{Scenario} \textbf{ AC2}: The scenario is about protrusion failure of the annular's sealing spherical element due to the passage of drill pipes and their joints. The annular element protrudes into the wellbore area. The test drill pipe joints can wedge it because it is made of an elastomeric material (rubber). Failures were discovered due to the impossibility of passing drill pipes and their joints or testing tools through the annular element and pressure tests performed during operation (subsea). The detection of the failure through the impossibility of passing drill pipe joints with the annular in operation was more common than through pressure tests in the records highly associated with this scenario. The corrective action for this scenario is the replacement of the component during maintenance.

\textbf{Scenario} \textbf{ AC3}: This scenario involves annular's piston damage such as pitting and scoring (the terms, scuffing, and scratches were used as synonyms too). Pitting failure is a type of corrosion that results in oxidation marks on the component. The scoring is usually found on the upper section of the piston, where it comes into contact with the adapter ring seals. According to some highly associated records, the scoring occurrence is related to piston damage that resulted in a raised edge between the piston and adapter ring. Other records indicated that foreign material was found on the chamber head. The majority of the failures were observed during routine maintenance operations through surface inspection with annular disassembly. The corrective action is the replacement of the component during maintenance. The drill's rotation generates acceptable metallic debris during drilling operations, and there is also ferrous debris leftover from milling or cutting the casing. When drilling mud exits the well and returns to the surface, rock debris, metallic and ferrous particles pass through the BOP stack, potentially entering the cavities of the annular and shear ram preventers and damaging parts such as the piston. A few records related to this scenario indicated that the cause of failure was that there is nothing to protect the piston from scratches and also that the material removed from scratches was found at the top when the piston was returning to the open position. 

\textbf{Scenario} \textbf{ AC4}: This scenario is about annular's leakage failure caused by rolling seals. A leak path is formed when the seal rolls out of the seal profile. According to failure records, this failure occurs more frequently in the annular cap seal. While some failure records argue that this failure is a recurring issue with the cap seal and a design flaw, others claim that failures occurred due to maintenance errors, such as installing the incorrect seal. The seal rolling identification by disassembling and inspecting the annular after one pressure testing with leaks appears frequent in the highly associated failure records. Leakage frequently occurs through a weep hole, which is a hole placed downstream of a seal to open a leakage path and accuse the seal of losing integrity. The component replacement is the corrective action for this scenario.

\textbf{Scenario} \textbf{ SRC1}: This scenario involves leakage failures caused by damaged seals, such as o-rings, on the blind shear ram. Seal damage and wear can result in leakage paths and, as a result, fluid leakage. Failures were detected using functional tests and pressure tests, the majority of which were performed at the surface and the stump test of the blind shear ram. After visually detecting the leak, a common procedure is to bleed off the pressure and remove the BSR door from the body, and then damaged seals, such as pinched, and leakage paths are observed. Some highly related records show that the problem occurred more frequently with backup o-rings and polypak seals, but other records show that the failure can affect other seals. According to one of these maintenance records, the bonnet studs were coated with grit and dirt around the cylinder head, with even more debris on the inner piston. The corrective action for this scenario is the component replacement. Another common procedure appears to be the execution of chamber tests to verify the BSR's capability to perform its required function. The leak was more common in the blind shear ram door or door hinge than in others parts.\textbf{\space }

\textbf{Scenario} \textbf{ SRC2}: The fractures or cracks in the blind shear ram blade's bolts are the subjects of this scenario. Failures were detected by non-destructive tests and inspection of the component. According to the highly associated records, this failure is an ongoing issue, and it probably is a design or material issue. The corrective action is the replacement of the cracked blade bolts. The presence of the term end of well indicates that this failure scenario is typically detected at the ending of drilling operations procedures, such as surface tests.

\textbf{Scenario} \textbf{ SRC3}: This scenario is about the failure of blind shear ram's seals, which affects the component's capability to maintain the pressure levels required. High-pressure tests executed with the BOP system in the rig detected the failures. A common procedure after detecting a failure is to bleed off the pressure and disassemble the BSR to investigate the cause. According to the records, the failure was detected only during high-pressure tests, not during low-pressure tests that yielded successful results. In this scenario, the corrective action is to replace the seals in the BSR. According to some records, this failure was caused by a design flaw or a manufacturing error, while others report it was caused by seal wear and tear.

\textbf{Scenario} \textbf{ SRC4}: The leakage failure in the blind shear ram's doors, bonnets, and hinges is the subject of this scenario. Some highly related records indicated that these failures occurred due to problems with parts such as bolts and seals. The records reported that the occurrence was caused by wear in the components and parts, but it is worth noting that a few pointed to maintenance errors as the root cause of the failure. The majority of the failures were identified during routine maintenance operations by performing a chamber test on the BSR. In this scenario, the corrective action is to replace the damaged components or parts during maintenance. Another common procedure appears to be the execution of chamber tests to determine whether or not the component's leakage has been resolved.

\begin{table*}[!htbp]
\caption{{BOP Control System components related concepts and their highest loading terms} }
\label{tw-a70108a235cc}
\def\arraystretch{1}
\ignorespaces 
\centering 
\begin{tabulary}{\linewidth}{p{\dimexpr.10080000000000002\linewidth-2\tabcolsep}p{\dimexpr.10960000000000006\linewidth-2\tabcolsep}p{\dimexpr.7896\linewidth-2\tabcolsep}}
\tbltoprule Concept & Singular Values & Terms that contribute the most to each concept\\
\tblmidrule 
\multicolumn{3}{p{\dimexpr(1\linewidth-2\tabcolsep)}}{\textbf{Regulator's failures concepts}}\\
RC1 &
  7.379 &
  regulator; leak; pressure; vent; note; maintenance; bop; blue\_POD; instal; yellow\_POD; soak\_test; test; remove; fluid; o\_ring; replace; description; valve; POD; new; observe; rebuilt; seal; oem.\\
RC2 &
  3.358 &
  pressure; maintenance; description; POD; increase; function\_test; regulate; decrease; overhaul; surface; swap; subsea; maintain; yellow\_POD; pilot\_pressure; fail; root\_cause; set; check; spare; new; supply\_regulator; circuit; rov.\\
RC3 &
  3.092  &
  soak\_test; leak; root\_cause; replace\_new; blue\_POD; oem; maintenance; note; description; assembly; manifold\_regulator; yellow\_POD; supply\_regulator; function; spring\_housing; BOP; manual\_regulator; vent\_tube; function\_test; crew; spare; detect; start; weep\_hole; control\_POD.\\
RC4 &
  2.945 &
  valve; vent; constantly; vent\_tube; manifold\_regulator; pressure; soak\_test; time; replace\_new; function; root\_cause; drip; pilot\_regulator; stop; notice; blue\_POD; cycle; increase; range.\\
RC5 &
  2.755 &
  note; pressure; yellow\_POD; damage; manifold\_regulator; o\_ring; shear\_seal; vent\_port; seal\_plate; crew; blue\_POD; soak\_test; replace\_new; come; annular\_regulator; internal; BOP; inspect; subsea; low; revision; seal; component; regulate.\\
\multicolumn{3}{p{\dimexpr(1\linewidth-2\tabcolsep)}}{\textbf{Compensated chamber solenoid valve's failures concepts}}\\
SVC1 &
  6.738 &
  CCSV; leak; valve; solenoid; POD; function; blue; maintenance; note; soak\_test; remove; BOP; test; yellow\_POD; replace; fluid; instal; description; observe; open; inspect; vent; new; rebuilt; o\_ring.\\
SVC2 &
  3.999 &
  valve; test; maintenance; pass; built; built\_incorrectly; event; receive\_oem\_unused; pressure; straight; sustain; receive\_oem; dedicate; end\_of\_well; test\_bench; oem; apply; prior; warehouse; seal; vent; description; root\_cause; vent; fluid; damage.\\
SVC3 &
  3.578 &
  solenoid; shear\_seal; seal\_plate; common; troubleshoot; leak; BOP; replace\_oem; multiple; disassemble; bench\_test; test; solenoid\_bank; reactivation; report; cameron; fail; currently; rig; run; investigation; drain; wear; receive; surface.\\
\tblbottomrule 
\end{tabulary}\par 
\end{table*}
\textbf{Scenario} \textbf{ RC1}: This scenario is about leakage failure in the HKR and MKR regulators due to damaged seals, such as o-rings. The majority of the failures were detected during surface soak tests. After visually detecting a leak coming from the vent tube, a common procedure is to bleed off the pressure and disassemble the regulator, where damaged seals and leakage paths were discovered. The corrective action for this scenario is to replace or rebuild the damaged component during maintenance using a repair kit. Finally, the regulator is tested, and if no leaks are found, the maintenance is completed. According to the strongly related records, the failures were caused by wear and tear in the regulators and their parts.

\textbf{Scenario} \textbf{ RC2}: This scenario describes a failure caused by the HKR regulators' inability to reach and (or) maintain the required pressure, despite increase and decrease commands executed on the control panels. Highly related failure records indicated that pressure fluctuations above the required set pressure occurred. Failures were detected~during surface functional tests as well as with the BOP in subsea operation. According to the records, after the pressure oscillations were detected, the regulator was disassembled to investigate possible causes of the failure. The corrective action for this scenario is the replacement or the repair (overhaul) of the regulator during maintenance. Wear and tear in components, such as o-rings, was suspected as the cause of the failures. 

\textbf{Scenario} \textbf{ RC3}: This scenario involves regulator's leakage failure in both the HKR and the MKR. Failures were detected during functional tests at the surface and soak tests, with fluid leaking through the vent tube. This scenario is similar to scenario RC1. These leaks were detected by passing fluid through the weep hole, which is a hole placed downstream of a seal to open a leak path and accuse the spring housing seal of failing. The vent hole and vent tube are connected to the spring housing component, and also any damage to the seals, such as o-rings, can result in fluid entering the spring housing and being detected as a leak when flowing through the spring house seal's weep hole. The corrective action for this scenario is the replacement of the damaged components, such as o-rings and spring housing seals, during maintenance.

\textbf{Scenario} \textbf{ RC4}: This scenario is similar to the scenario RC3. The differences between the scenarios are due to the characteristics of the leakage. The vast majority of the cases involved drips. Soak tests and surface function tests were used to detect the failures. The component replacement during maintenance was the corrective action for this scenario.

\textbf{Scenario} \textbf{ RC5}: This scenario describes HKR regulator leakage caused by damaged components such as shear seals, o-rings, and seal plates. The failures were detected during surface soak tests, with the leaking occurring through the vent port. Following the detection, the regulator was disassembled in order to determine the cause of the failure. According to some highly associated records, they found damages in the shear seals, such as cut, extruding, and nibbling o-rings, shear seals, and scratches in the seal plate. The corrective action for this scenario is to replace damaged components during maintenance.

\textbf{Scenario} \textbf{ SVC1}: The failure of compensated chamber solenoid valves due to leakage is the subject of this scenario. Soak tests and functional tests, both performed at the surface, were used to detect the failures. According to highly related records, leakage frequently occurs in the vent tube. After visually detecting a leak from the vent tube, a common procedure is to remove and disassemble the CCSV for investigation. The immediate corrective action for this scenario is to replace the component, which can be a rebuilt or new valve, and test to see if the failure has been resolved. The records with a high score in this scenario indicated a variety of failure causes. They report that the leaks were caused by wear and tear on the seal plates and damage to the seals, including the seal assembly, o-rings, and shear seals. Some of the seal plates had been worn and scuffed.

\textbf{Scenario} \textbf{ SVC2}: This scenario is similar to scenario SVC1. According to the highly related records, when pressure was applied, the valve would not seal and allow fluid to pass straight through to the vent. The tests were conducted at the surface, probably near the well's end or before the event.

\textbf{Scenario} \textbf{ SVC3}: This scenario is about CCSV's leakage. Leaks were discovered during functional tests performed at the surface due to a stream of fluid passing through the common vent or common drain on the solenoid bank. According to the highly associated records, a common procedure is troubleshooting to identify the specific compensated chamber solenoid valve leaking and then disassembling to investigate. Some reported finding scratches and wear signals on the seal plate and shear seal and leaks caused by wear and tear. In this scenario, the corrective action is to replace the entire CCSV or just the seals during maintenance.
    
\section{Discussion}
The failure scenarios reported in this research evidenced that critical failure modes in BOP system's components can occur for various reasons and can be detected using a variety of methods. Some scenarios were similar to others, but they differed in minor ways, such as the failures observed and detection methods or detection conditions. For example, some annular leakage failures occurred due to different component problems, such as damaged head seals, chamber seals, or piston seals, and these failures were sometimes detected through pressure tests and other times through functional tests. Despite the use of the singular values elbow graph, domain experts were essential to solve the problem of analyzing failure scenarios that were very similar to others and concepts that lacked essential parameters, such as failure modes, by helping to determine the cut-off number of concepts to be considered for analysis in this study. It is worth noting that some of the excluded concepts had similar terms and high loading failure records to the selected concepts for analysis. They may represent a variation of the failure scenario, taking into account a less frequent dimension, such as one different detection method.

LSA results highlighted the scenarios of failure that were more frequent in the RAPID-S53 non-planned corrective maintenance records of the BOP system.  Three or more failure scenarios reported more frequently in the database were found for the components under consideration, as expected by the authors. Also was possible to identify the affected systems, subsystems, components, detection methods used, observed failures, and immediate corrective actions taken to solve the problem for all scenarios. Aside from the fact that the components interact with each other,  findings show that failures and maintenance actions differ depending on the component and even within the same component due to the type of failure observed. This characteristic of the results is most likely due to the complexity of the blowout preventer system and its components, which contain a large number of subcomponents and parts.

To fully comprehend the findings of this study, it is essential to assess the similarities and differences between the failure scenarios and the findings of other studies related to the selected BOP system's components. Annular failure scenarios reported leakage failures caused by damaged or rolled seals, annular's piston damage, and protrusion of the annular's sealing spherical element. In\unskip~\cite{1143479:22575643}, was observed that six of the 12 annular preventer's failures collected in the study were internal leakage (leakage through a closed annular) failures, while the other six were failures to fully open. In a more recent study, out of a total of 24 annular preventer's failures collected, failed to fully open and internal leakage were the most common failure modes, accounting for 15 of the failures\unskip~\cite{1143479:22575614}. These observed major failure modes are present in the failure scenarios. The internal leakage failure mode is included in the leakage failure, and the protrusion of the annular's sealing spherical element, which is part of scenario AC2, is equivalent to the failure to fully open. One internal leakage failure collected in\unskip~\cite{1143479:22575614} had a similar context to scenario AC3 because after opening the upper annular preventer, some metal swarf was observed stuck between the annular piston and the housing, and the corrective action was to remove the piston, adaptor ring, and packing element and replace all with new parts. In\unskip~\cite{1143479:22575686}, many internal leakage failures that occurred through the weep holes due to damaged seals, as seen in scenarios AC1 and AC4, were reported. They also reported one undetailed annular failure discovered during testing before running the BOP, which was corrected by replacing the piston. Similar to the failures in scenario AC3, this failure was probably discovered during inspections of the disassembled annular preventer.

The detection method varies depending on the failure observed, but in general, in annular preventer's failure scenarios, surface pressure tests and subsea pressure tests are the most frequent, followed by inspection and surface functional tests. Several failure records associated with the concepts reported using hydraulic tests in conjunction with the disassembled annular inspection to identify the causes of leakage failures, a fact which\unskip~\cite{1143479:22575686} also observed. Most fully open failures observed in reliability studies were due to the inability to pass drill pipes or testing tools through the annular spherical element, necessitating an increase in tool load (over-pull), and leakage failures are primarily identified through subsea pressure tests\unskip~\cite{1143479:22575686,1143479:22575614}. Furthermore, the most common maintenance corrective actions for annular preventer's failure scenarios were component replacement, but in some cases, the entire annular preventer was replaced. These maintenance procedures were observed for failures collected in other reliability studies from 1999 to 2019\unskip~\cite{1143479:22575643,1143479:22575686,1143479:22575614}, suggesting that having a standby annular preventer stored is less expensive than the non-productive time costs of awaiting repair affected components in some cases.

The shear rams' failures included both blind shear ram's and casing shear ram's failures, but the high loading terms and records for each concept only included BSR's failures. CSR's failures were almost certainly reported much less frequently than BSR's failures, and this could be examined further in a more focused study. When the pressure within the drilling system becomes uncontrollable, the blind shear ram in a BOP system is the critical last line of defense, and if the BSR is available on-demand, a blowout will not occur\unskip~\cite{1143479:22575645}. Thus, the BSR is a critical component to the safety of drilling operations and the blowout preventer system. Failure modes of leakage due to damaged seals, damaged components such as doors and hinges, fractures and crackings in the bolts, and inability to maintain required pressure levels due to damaged seals were all present in the blind shear rams' scenarios. The two previous reliability studies\unskip~\cite{1143479:22575686,1143479:22575614} collected and analyzed a few blind shear ram's failures, and when the two studies were combined, a total of six blind shear ram's failures were gathered. Five of the six failures were caused by leakages in the BSRs, four of which were internal leakages (through a closed ram), and one was an external leakage. One internal leakage was reported without the affected component or the causes specified, so this failure could be included in scenarios SRC1, SRC3, or SRC4, depending on the real unreported causes and detection methods. Another internal leakage occurred due to lateral t-seal failure, which allows for the existence of leakage paths and is compatible with scenario SRC1. A failure in the bonnet seals caused one other internal leakage, and the external leakage was caused by a failure in the bonnet door seals, both of which are similar to the highly related failures observed in scenario SRC4. The fifth reported leakage failure was leakage through the BSR EVO lock motor while it was in the unlock position. The sixth failure occurred when a BSR failed to close due to the hydraulic hose not being connected to the manifold, and the failure occurred during BOP's maintenance. None of the previously reported failures are compatible with SRC2, which involves fractures or cracks in the shear ram blade' bolts.

Surface pressure tests are the most frequently reported method of detecting BSR's failures in the blind shear ram's preventer failure scenarios, followed by inspection and surface functional tests. The use of hydraulic tests to identify leaks, followed by disassembling the blind shear ram's for inspection to determine the causes of the failures, is a common procedure in the scenarios' highly related failures. Three of the five leakage failures identified in the previous reliability studies\unskip~\cite{1143479:22575614,1143479:22575686} were detected during the BOP installation testing, one through pressure tests and one during the execution of a function test. In three of the leakage failures, the blowout preventer was pulled to repair and replace the affected components, one did not specify the corrective actions, and one reported that the failure was corrected thirteen days later when the BOP was on the rig for other reasons. The scenarios indicated more failures in tests performed with the blowout preventer in the rig and did not address the need to pull the BOP from subsea to the rig. However, the failures reported in the reliability studies, in which the BOP was pulled, and the drilling company lost productive time, highlight the importance of the blind shear ram preventer to drilling operations safety.

This study's regulator's failure scenarios indicated leakage due to damaged seals, component's failures such as seal plates and spring housing, and failures due to the regulator's inability to reach and (or) maintain the required pressure. The findings, such as scenarios RC1 and RC2 (which suggested seal wear and tear as the cause of failure), support some scholars' perception\unskip~\cite{1143479:22575936} that were deteriorating shear seals (i.e., through scoring, erosion) one of the primary causes of fluid leakages in the regulator component because a deteriorated seal on either the supply or vent spool cannot isolate the connecting paths. This fact strengthens the link perceived in the scenarios between fluid leakage and damaged seals. Regulator's internal leaks are an important failure mode to the BOP system safety because they are undesirable states that can impact the system's expected life, operational availability, and overall system and environmental safety\unskip~\cite{1143479:22575936}. The scenarios RC3, RC4, and RC5 indicated leakages that can occur because of a deteriorated seal, but the scenarios are compatible with leakages caused by damages as cut, extruding, and nibbling seals. The scenario RC2 represents a potentially dangerous failure of the BOP system because pressure oscillations created or not damped by a pressure regulator cause dynamic loading of neighboring components that may not have been considered during their design\unskip~\cite{1143479:22575936}. In some cases, the scenarios RC1, RC3, and RC5 can cause pressure oscillations and the regulator's inability to maintain the required pressure, as indicated in RC2, but this depends on the leakage characteristics.

Compensated chamber solenoid valves are an essential component of the BOP control system, and the failure scenarios SVC1, SVC2, and SVC3 indicated that leakages were the most frequently observed failures, with the differences between the scenarios relying mainly on maintenance operations. When an operator commands the BOP system to perform a specific function, a signal is sent from the central control unit to the SEM for decoding and then to a specific compensated chamber solenoid valve (placed in the POD and associated with the desired function) that opens, causing an SPM valve to change position and allowing high-pressure fluid stored in the accumulator to flow\unskip~\cite{1143479:22575730}. Because each function has its own compensated chamber solenoid valve and some functions are more critical to the BOP system than others (e.g., blind shear ram close, annular preventer close), if the CCSV fails to trigger the subplate-mounted valve, the function affected defines the risk associated with the failure. Failures in compensated chamber solenoid valves may result in an inability to command the other components studied and, as a result, perform their required functions. Leaks in control system components, such as solenoid valves and regulators, increase operating costs, while excessive leakage can lead to a loss of hydraulic pressure in the control circuit\unskip~\cite{1143479:22575855}.

The majority of failures in regulator's failure scenarios were detected via surface functional tests or inspection, whereas the main detection methods in CCSV scenarios were functional tests and soak tests, both of which were performed with the BOP in the rig. The procedure to inspect the component to investigate the cause of failure after the leakage is detected through hydraulic tests is also common in the highly related failures of compensated chamber solenoid valve and regulator's scenarios. Thus, the scenarios of the four components point to this as a common procedure in the oil and gas industry, at least for failures of the components under consideration.

The findings indicated that tests were critical for detecting failures of the four components studied and for the safety of the drilling phase operations and activities, as tests procedures appear in the scenarios and high related failure records of the DC matrix. Even though the textual failure records in RAPID-S53 are generated by a wide range of drilling contractors, each of which has its monitoring procedures, the scenarios indicated specific procedures capable of detecting different failure modes. These procedures are likely to be more effective in monitoring the capability of studied components to perform their required functions. However, many failures were detected during tests performed with the blowout preventer in the rig, which presents a challenge for the companies because these tests are expensive to monitor the BOP system condition. When the blowout preventer is pulled to the rig, drilling operations should be halted, and the companies must bear the costs of retrieving and maintaining the BOP system and non-productive time for the entire rig. Another important question raised by the scenarios is the effectiveness of the tests or other monitoring procedures conducted with the blowout preventer in subsea to monitor the capability of the components studied in performing their required functions, and why the majority of failures were detected by tests executed while the system was on the rig. If the pressure within the drilling system becomes uncontrollable, and the BOP is subsea with components such as the annular preventer and shear ram preventer unavailable to close the wellbore on demand, the entire rig is in danger.
    
\section{Conclusion}
This paper uses a text mining approach based on the well-known LSA method to extract information and understand the main failure scenarios of four blowout preventer's components: annular preventers, shear ram preventers, regulators, and compensated chamber solenoid valves. A total of 1312 failures and corrective maintenance descriptions reported by engineers and technicians in charge of BOP system maintenance is examined in this research to assess the most common failures, detection methods, repair practices, causes of failures, challenges that drilling companies face, and their experience with BOP system's failures for each component. At least three main failure scenarios were found for each of the four BOP's components studied based on unstructured failure descriptions. Although this study only examined failure descriptions for four components, it represents a step toward understanding the BOP system's main failure scenarios.

This research has both theoretical and practical potentialities. In an academic sense, unlike previous research, this study focused solely on textual descriptions of the failures, which contained information such as failure mode, causes, and detection method, and contributes to an understanding of the main failures of the BOP system. Given the scarcity of research on the most common blowout preventer failures and maintenance procedures, the failure scenarios presented in this paper may be useful for future studies. There is also a lack of information available on BOP test procedures that were more effective in detecting certain types of failures. Scholars may use the scenarios to examine the maintenance procedures associated with corrective maintenance, to support reliability and safety analysis, and risk management with information about the failures, or as an input to develop theoretical models of BOP system condition monitoring for the components studied. 

Furthermore, the findings indicating tests performed with the blowout preventer in the rig as a main detection method of the failures require extra attention since this reinforces the importance of the recent efforts of scholars, agencies, and oil and gas companies to increase the reliability of the BOP system. Few studies used text mining to analyze maintenance descriptions (texts that are very technical and complex), and the approach described in this paper has the potential to mitigate the technical vocabulary problem inherent in maintenance data. It produced promising results when analyzing unstructured failure descriptions to identify the main failure scenarios. Through customization and adaptation, the approach presented in this study can be used to extract relevant information of maintenance procedures contained in databases of others equipment.

Drilling companies and BOP manufacturers can use the failure scenarios to improve their services and products. Even though each drilling company has its procedures for investigating anomalies detected in the BOP system during drilling activities, managers in charge of blowout preventer systems can use the scenarios to support decisions when investigating anomalies on the components studied because they indicated detection methods that are likely more effective in detecting specific failures. Understanding the most common failure modes, detection procedures capable of identifying them, and their causes can also support engineers and technicians to develop better maintenance procedures and reliability analysis of the blowout preventer system. The findings establish a level of relevance (the singular values) between each component's failure scenarios, allowing drilling contractors to focus their efforts on monitoring the failures that occur more frequently. The level of relevance is critical to reducing the costs of the activities carried out during the drilling phase. The failure scenarios can provide a reliable assessment of component failures to BOP manufacturers, which can aid in designing better products or services. Regulatory agencies, for example, can use the findings to encourage oil and gas companies to take action, which is critical for improving industry safety by lowering accident risks while increasing equipment availability.

The findings may prompt oil and gas companies to reconsider the testing procedures used on BOP system's components. The testing of the blowout preventer system is critical to increasing safety, as demonstrated by the scenarios. However, the more frequently the BOP is tested, the lower the system's availability. Testing the system is detrimental not only to its availability but also because each function test, even if it does not overload the system, accounts for one less function in the system's useful life. If a component is designed to be used 200 times, it will lose 1/200 of its useful life for each function test. When examining components over a long period, function tests can significantly reduce their expected service life. For example, the service life of an annular preventer is usually 500 pressurizations or 15 years for its top cover and housing\unskip~\cite{1143479:22575711,1143479:22576316}, so if the component is tested at least once a week, assuming that it was not pressurized for anything except tests, the annular will have a service life of about ten years. The annular's estimated service life is five years if one pressurization per week due to drilling activities is included, three years if two pressurizations are included, and so on. This effect is worsened when investigating the impact of pressure testing, as it overloads the system. Therefore, drilling companies must consider the effectiveness of the procedures they use to investigate anomalies detected in the BOP system because they can reduce the service life of the components to a fraction of what it should be. As previously stated, the failure scenarios can assist them in determining which procedures are likely to be more effective.
    
\section{Limitations}
Although this study is a step forward in using failure records textual data, it is worth noting some limitations. The non-planned corrective maintenance descriptions were added to the RAPID-S53 through an open question on the failure event registration form, and the descriptions contain some subjectivity, even though the text is highly technical. There is a difference in the quality of the records when considering the volume of information in the descriptions, which varies according to experience and engagement in reporting all the details of the failures. The entire investigation has been reported in some descriptions, including details such as failure mechanisms, proper as input to discussions about condition-based maintenance. However, there are others cases with very brief descriptions of corrective maintenance, which is detrimental to the database and research. These two types of records are few and have little impact on LSA results, but drilling companies must strive to improve the engagement of the workers responsible for reporting failures so that the descriptions contain more information than they have now.

Based on a well-known text mining technique, the analysis generated knowledge about the most common failures and corrective maintenance procedures related to the BOP system's components studied. On the other hand, because the failure records were only collected from the RAPID-S53, the results obtained are restricted to a single database. Besides that, even though the database have reports from regions across all globe, the majority of failure records are from drilling companies working in the United States part of the Gulf of Mexico, and because of that, the results of the analysis may not reflect the main failure modes in all countries. While failure descriptions for annular preventers and shear ram preventers in the US part of the Gulf of Mexico correspond to approximately 60-65 percent of the data examined, failure descriptions for regulators and compensated chamber solenoid valves in the same location correspond to approximately 80 percent.

The scenarios were presented so that the LSA generates concepts (from higher to lower singular values), which can be related to the frequency of occurrence in the textual failure records. Some failure scenarios that appear as second, third, or fourth may pose a greater risk to the rig's safety than failures indicated in the first scenario of the components. Thus, the severity of the failure modes is not taken into account by the latent semantic analysis. LSA is carried out through systematic and mathematical analysis, but there is human involvement in interpreting the concepts, which introduces subjectivity. The relevance of analyst knowledge to the outcome of text mining using the LSA approach cannot be underestimated. This limitation was addressed through a revision conducted with a team of researchers, engineers, and technical workers with oil and gas industry experience.

\section{Suggestions}
The study's findings shed light on the breakdown of failure events by components' failures modes,\textbf{\space }detection methods, corrective actions, suspected causes, and the frequency with which they occur. Future research could use the results of a scientifically based procedure as a base for reliability studies in the BOP system's components under consideration, as well as the development of theoretical models of condition monitoring, maintenance procedure assessment, risk management, and safety analysis. The industry regulations and standard requirements have been developed and evolved to improve the safety of the drilling phase\unskip~\cite{1143479:22575742,1143479:22575716}. However, the findings reinforce that oil and gas companies, regulatory agencies, technology institutes, and scholars must research intelligent and practical solutions to improve BOP system availability while maintaining or improving safety. It is critical for the future of drilling oil and gas companies to increase the effectiveness of monitoring procedures by pursuing innovative ways to monitor the BOP system, such as new technologies, techniques, and approaches related to condition-based maintenance of the components as other intelligent solutions. These efforts are essential to increase the availability of the blowout preventer system, and the safety of the rig crew and the environment. 

In\unskip~\cite{1143479:22575645}, for example, a state-based approach is proposed to analyze unavailability by incorporating BSR preventer testing activities into a multiphase Markov process, and investigated the effects of testing errors and delayed repairs on the likelihood of component unavailability. In\unskip~\cite{1143479:22575727}, an integrated a fault tree analysis updated with a near real-time failure database is used into the operational decision-making process to reduce non-productive time on drilling rigs due to complex failure propagation within the BOP system. In\unskip~\cite{1143479:22575711}, a high-fidelity method is proposed to monitoring the stress of the annular preventer's top cover and housing that combines the theoretical calculation method (TCM), finite-element analysis (FEA), and stress testing experiment (STE). In\unskip~\cite{1143479:22575740}, an adaptive model-based approach is proposed to real-time condition monitoring of annular preventer's functions by combining a first-principles model with in-field data by adapting model coefficients, interpreted as annular preventer's health indicators. These are examples of recent studies that presented new solutions for improving drilling operations' safety while increasing the availability of BOP system's components, both of which are critical to the oil and gas industry's future and sustainability. Efforts in this direction are likely to improve maintenance management and industry safety by reducing accident risks.

The discussion in this study can be expanded by using cuttings from the database to other important components of the blowout preventer system, such as pipe rams and SPM valves, to understand their main failure scenarios. Furthermore, maintenance intelligent solutions, such as the gathering of video reports, have potential to increase the volume of information described by technicians and engineers. In this context, the challenge of a human expert examining each service entry can rise substantially. So, the ability of text mining methods, such as the LSA, to reduce the dimensions of large volumes of data to a more manageable number without losing a significant amount of information can be explored by scholars and oil and gas companies to extract useful information from textual descriptions of maintenance procedures.
\section*{Acknowledgements}The authors wish to acknowledge the financial support of Brazil's national oil company, Petrobr\'{a}s (No.ANP: 20741-5).

\bibliographystyle{elsarticle-num}

\bibliography{article.bib}

\end{document}